\documentclass[lettersize,journal]{IEEEtran}

\usepackage{array}
\usepackage[caption=false,font=normalsize,labelfont=sf,textfont=sf]{subfig}
\usepackage{stfloats}
\usepackage{cite}
\usepackage{url}
\usepackage{textcomp}
\usepackage{xcolor}
\usepackage{hyperref}
\usepackage{booktabs}	% 插表格用的宏包
\usepackage{diagbox}    % 插表格用的宏包
\usepackage{multirow}   % 插多行表格用的宏包
\usepackage{verbatim}	% 多行注释用的宏包
\usepackage{graphicx}
\usepackage{tabularx}

\usepackage{amsmath,amssymb,amsfonts,graphicx}
\usepackage{epstopdf}

\newtheorem{definition}{Definition}[section]

\newtheorem{property}{Property}[section]
\newtheorem{strategy}{Strategy}    % gws
\usepackage{graphicx,epstopdf,algpseudocode,caption,url}   % gws
\usepackage{graphicx}
\usepackage{caption}
\usepackage{booktabs}
\newenvironment{proof}{\begin{IEEEproof}}{\end{IEEEproof}}  % gws
\usepackage[ruled,linesnumbered]{algorithm2e}
\usepackage{amssymb}
\usepackage{hyperref}
\usepackage{bm}
\usepackage[switch]{lineno} % 里面的选项代表双栏

\begin{document}

\title{Targeted Mining Precise-positioning Episode Rules}

\author{Jian Zhu,~\IEEEmembership{Member,~IEEE}, Xiaoye Chen, Wensheng Gan,~\IEEEmembership{Member,~IEEE},\\ Zefeng Chen, Philip S. Yu,~\IEEEmembership{Life Fellow,~IEEE}

\thanks{This research was supported in part by the National Natural Science Foundation of China (Nos. 62237001, 62272196, 62122022), National Key R\&D Program of China (No. 2021ZD0111501), Guangdong Basic and Applied Basic Research Foundation (Nos. 2022A1515011590, 2022A1515011861), Guangzhou Basic and Applied Basic Research Foundation (No. 2024A04J9971). (Corresponding author: Wensheng Gan)}
    
\thanks{Jian Zhu and Xiaoye Chen are with the School of Computer Science and Technology, Guangdong University of Technology, Guangzhou 510006, China. (E-mail: dr.zhuj@gmail.com, cxx\_3727@outlook.com)}

\thanks{Wensheng Gan and Zefeng Chen are with the College of Cyber Security, Jinan University, Guangzhou 510632, China. (E-mail: wsgan001@gmail.com, czf1027@gmail.com)}
	
\thanks{Philip S. Yu is with the University of Illinois Chicago, Chicago, USA. (E-mail: psyu@uic.edu)} 
}

% The paper headers
\markboth{IEEE Transactions on Emerging Topics in Computational Intelligence~2024}{}

\maketitle

\begin{abstract}
    The era characterized by an exponential increase in data has led to the widespread adoption of data intelligence as a crucial task. Within the field of data mining, frequent episode mining has emerged as an effective tool for extracting valuable and essential information from event sequences. Various algorithms have been developed to discover frequent episodes and subsequently derive episode rules using the frequency function and anti-monotonicity principles. However, currently, there is a lack of algorithms specifically designed for mining episode rules that encompass user-specified query episodes. To address this challenge and enable the mining of target episode rules, we introduce the definition of targeted precise-positioning episode rules and formulate the problem of targeted mining precise-positioning episode rules. Most importantly, we develop an algorithm called Targeted Mining Precision Episode Rules (TaMIPER) to address the problem and optimize it using four proposed strategies, leading to significant reductions in both time and space resource requirements. As a result, TaMIPER offers high accuracy and efficiency in mining episode rules of user interest and holds promising potential for prediction tasks in various domains, such as weather observation, network intrusion, and e-commerce. Experimental results on six real datasets demonstrate the exceptional performance of TaMIPER.
\end{abstract}


\section{Introduction}  \label{sec:introduction}

\IEEEPARstart{F}{requent} episode mining (FEM) \cite{fournier2022pattern,ouarem2024survey} is a widely recognized task in the field of data mining that aims to identify all frequent episodes within a single time series with a frequency surpassing a specified threshold \cite{ao2017mining, ouarem2021mining}. FEM tasks entail the consideration of two distinct types of input sequences: simple sequences, where each event corresponds to a unique timestamp, and complex sequences, where multiple events can occur simultaneously \cite{fournier2020tke}. The versatility of FEM allows it to be applied across a broad spectrum of scenarios relating to time series data, including traffic data \cite{ao2017mining}, web navigation logs \cite{casas2003discovering}, financial data \cite{ng2003mining}, event detection in sensor networks \cite{wan2009frequent}, etc. Finding frequent episode rules (FERs) from a set of frequent episodes is a fundamental problem in FEM \cite{ao2017mining}. FERs, which are typically expressed as `$\alpha \rightarrow \beta$', represent relationships in which the occurrence of the `$\beta$' episode follows that of the `$\alpha$' episode. Determining FERs involves utilizing frequent events as antecedents, which must surpass a user-defined confidence threshold in terms of occurrence. The discovery of FERs serves to enhance the understanding, analysis, and interpretation of time series data, enabling predictions and informed decision-making.

\begin{figure*}[ht]
    \centering
    \includegraphics[clip,scale=0.68]{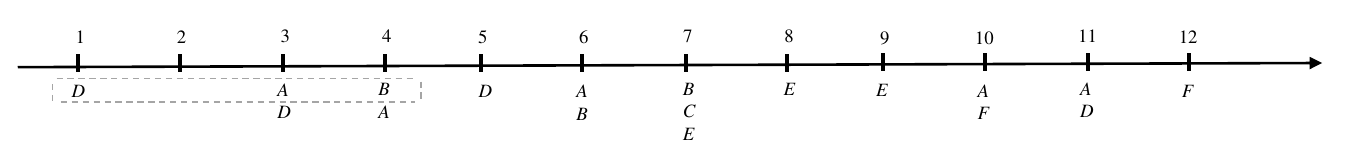}
    \caption{A sequence of events. This sequence denotes the presence of event $D$ at moment 1, the absence of any event at moment 2, the co-occurrence of events $A$ and $D$ at moment 3, the co-occurrence of events $A$ and $B$ at moment 4, and so on for subsequent moments.}
    \label{fig:events}
\end{figure*}

Fig. \ref{fig:events} displays an example of simulated data for illustrating a series of events. By leveraging the information depicted in Fig. \ref{fig:events}, we proceed to provide a more in-depth introduction to FER. Disregarding frequency considerations, we can derive a FER: $<$$D$$>$ $\rightarrow$ $<$$A$, $B$$>$. Here, events $D$, $A$, and $B$ correspond to moments 1, 3, and 4, respectively. This rule implies that after the occurrence of event $D$, events $A$ and $B$ ensue. Event $D$ serves as the antecedent, while events $A$ and $B$ act as subsequent events in the rule. In general, FERs also need to take into account time constraints. Specifically, in `$\alpha \rightarrow \beta$', it is crucial to define the time delay between the occurrence of `$\alpha$' and `$\beta$' and how `$\beta$' unfolds. In existing FEM algorithms, most can only approximate the time range within which `$\beta$' occurs. However, certain algorithms \cite{ao2017mining} can mine the exact occurrence time of each event within `$\beta$', enabling precise determination of time intervals between events. Consequently, for the event sequence in Fig. \ref{fig:events}, the earlier introduced rule can be expressed as `$<$$D$$>$ $\stackrel{2}{\rightarrow}$ ($<$$A$, $B$$>$, $<$1$>$)' within the framework of precise positioning episode rules (PER) when considering time constraints. This rule indicates that `$\beta$' occurs after `$\alpha$' with a time interval of 2, and within `$\beta$', the event $B$ occurs one-time interval after the event $A$. PER provides precise timestamps, expanding the scope and effectiveness of its application. 

However, episode rule mining can be more targeted. This implies that episode rule mining can be imbued with a target, enabling purposeful exploration and significantly boosting efficiency. Current episode rule mining tasks do not emphasize user-defined mining targets, despite the wide range of potential valuable applications for targeted episode rules. For instance, within the domains of economics and finance, if the objective is to discern when stock prices rise after specific events, traditional FEM methods may produce an overwhelming number of rules, encompassing rising, falling, and stable prices alike. Such rules, often irrelevant to the primary objective, may even outnumber genuinely valuable ones. Similarly, in the field of biomedicine, if medical practitioners are interested in specific DNA sequences, but the mining results generate an inundation of unrelated rules, it can lead to a significant waste of resources and effort.

To address these challenges, this paper presents a novel approach inspired by a few previous algorithms \cite{ao2017mining, huang2024taspm}. Specifically, we define target rules, formulate the problem of targeted mining of precise episode rules, introduce four pruning strategies, and present a novel algorithm called Targeted Mining of Episode Rules (TaMIPER). The distinctive characteristic of TaMIPER lies in its capacity to accurately identify rules that satisfy user-defined targets while adhering to predetermined support and confidence thresholds. The rules generated by TaMIPER encompass antecedent events, consequent events, and time constraints. The antecedents are derived from minimal episodes, while the consequents are constructed using fixed-gap episodes and must encompass all target events specified by the user. The main contributions of this paper can be summarized as follows:

\begin{itemize}
    \item To address the needs of target-sensitive applications, we introduce the concept of precise target episode rule, define query episodes, and formalize related issues.
    
    \item We develop an efficient algorithm called TaMIPER, which can discover complete, accurate, and target-compliant episode rules. 
    
    \item We propose four pruning strategies that significantly reduce resource costs, such as time and space requirements associated with the mining process.
    
    \item We conduct experiments using various real datasets, and the results demonstrate the superior performance of TaMIPER compared to the baseline approaches of targeted episode rule mining.
\end{itemize}

The remaining sections of this paper are as follows. Section \ref{sec: relatedwork} discusses related work. Section \ref{sec: preliminaries} presents the preliminary definition and problem statement for targeted mining of precise episode rules. Section \ref{sec: algorithm} introduces our efficient algorithm, TaMIPER. Section \ref{sec: experiments} covers experimental evaluation and results. Finally, we conclude this paper and discuss future work in Section \ref{sec: conclusion}.

\section{Related Work} \label{sec: relatedwork}

\subsection{Constrained Pattern Mining}

Pattern mining \cite{tian2018surrogate, li2021frequent, wu2023opr}, a widely employed task in data mining and machine learning, is dedicated to the discovery of significant and recurring patterns (e.g., itemsets\cite{kubat2003itemset}, rules\cite{fister2021information}, and episodes\cite{ao2017mining}) within the datasets. Nowadays, there has been substantial progress in this fundamental field. In the realm of sequential pattern mining (SPM) \cite{gan2019survey}, several classic algorithms have been introduced, including SPADE \cite{zaki2001spade}, PrefixSpan \cite{pei2004mining}, and CloSpan \cite{yan2003clospan}. In structured pattern mining, algorithmic evolution has transitioned from simple to complex patterns, advancing towards trees, lattices, and graphs \cite{fournier2022pattern}. Pattern mining has demonstrated remarkable success in related areas, such as pattern clustering. For SPM, various tasks have emerged, addressing specific constraints, such as top-$k$ patterns \cite{fournier2013tks}, contiguous patterns \cite{zhang2015ccspan}, closed patterns \cite{bonchi2004closed}, gap-constrained patterns \cite{wu2017nosep}, uncertain patterns \cite{ahmed2020evolutionary, chen2024uucpm}, and patterns reflecting multiple constraints in SPM \cite{wan2022fast,zheng2023fast}. To accommodate variations in values for each item, the concept of high utility SPM \cite{gan2019huopm,gan2020survey}, along with corresponding algorithms, has been extensively researched and applied. Addressing the occurrence time constraint for each event in the results, Ao \textit{et al.} \cite{ao2017mining} introduced the concept of fixed gap sets and conducted mining based on this framework. Furthermore, some excellent algorithms, such as f-NSP \cite{dong2018f}, sc-NSP \cite{gao2023toward}, and NegI-NSP \cite{qiu2023efficient}, have been proposed to mine negative sequential patterns. 

\subsection{Frequent Episode Mining}

An episode is defined as a non-empty ordered sequence of events, and frequent episode mining (FEM) involves the identification of recurring episodes within a single sequence \cite{ouarem2024survey}. Pioneering algorithms in this task, WINEPI and MINEPI \cite{mannila1997discovery}, offer two approaches for discovering frequent episodes: sliding windows and minimum occurrences. However, the mining process may generate numerous unnecessary combinations, leaving room for improvement in terms of time and memory resource consumption. To address this, Huang and Chang \cite{huang2008efficient} modified the vertical-based MINEPI to MINEPI+ and introduced the EMMA algorithm, which utilizes memory anchoring to reduce the search space in mining episodes from complex sequences. Ma \textit{et al.} \cite{ma2004finding} enhanced the MINEPI algorithm's performance and proposed algorithms such as Position Pair Set (PPS), which employ prefix growth in the mining of frequent episodes without generating candidate items. Laxman \textit{et al.} \cite{laxman2005discovering} introduced a novel definition of episode frequency, namely the non-overlapping occurrences of interludes, and presented a new computational algorithm. Wu \textit{et al.} \cite{wu2013mining} integrated the concept of utility mining \cite{gan2020survey} into episode mining, and then proposed the UP-Span algorithm and the Episode-Weighted Utility (EWU) model, efficiently addressing the problem of utility-based episode mining by combining these approaches. The TKE algorithm \cite{fournier2020tke} addressed the \textit{minSup} setting problem by defining top-$k$ frequent sets.

\subsection{Episode Rule Mining}

Episode rule mining is a further exploration of frequent episodes, aiming to uncover robust relationships between events and analyze sequences of events. Traditional algorithms \cite{huang2008efficient, mannila1997discovery} are designed to mine all episode rules in a time database that meet the thresholds of \textit{minSup} and \textit{minConf}. WinMiner \cite{meger2004constraint} explores episode rules based on the maximum gap constraint and offers recommendations for determining the window size that maximizes local confidence. In response to various time-sensitive requirements in numerous applications, Ao \textit{et al.} \cite{ao2017mining} introduced the concept of a fixed gap set and mines precisely located episode rules. In a recent development, \cite{fournier2021mining} presents the POERM algorithm, an efficient partially ordered episode rule miner. The NONEPI algorithm \cite{ouarem2021mining} specializes in extracting episode rules with non-overlapping occurrences. To enhance the discovery of frequent episodes, Gan \textit{et al.} \cite{gan2022discovering} introduced the Episode-Weighted Utility (EWU) concept and proposed the utility-driven episode mining framework called UMEpi. Note that most of the current episode rule mining algorithms are based on frequency.

\subsection{Targeted Pattern Mining}

Targeted pattern mining (TaPM) has emerged as a solution to selectively query patterns aligned with user requirements, enabling the early avoidance of ineffective or unnecessary mining efforts. Kubat \textit{et al.} \cite{kubat2003itemset} designed a sequentially independent itemset tree to transform the database into a project tree that facilitates information access. However, this approach necessitates additional memory. To overcome this limitation and enhance the scalability of the itemset tree, the memory-efficient itemset tree (MEIT) \cite{fournier2013meit} can reduce the size of IT nodes through a node-compression mechanism. Building on the itemset tree, the Apriori principle can be applied to avoid generating uncommon itemsets, resulting in faster item generation and extension rule sets \cite{lewis2019enhancing}. Miao \textit{et al.} \cite{miao2022targeted} introduced the TargetUM algorithm, pioneering goal-oriented efficient itemset mining. To enhance operational efficiency on large-scale datasets and multiple sequence datasets, TaSPM \cite{huang2024taspm} employs bitmaps to search for target sequential patterns. TaSRM \cite{gan2022towards} is designed for targeted mining sequential rules. TUSQ \cite{zhang2022tusq} incorporates utility into target sequence query tasks, utilizing a data structure called the targeted chain and employing projection techniques to address query tasks. To introduce greater flexibility in gap constraints for targeted sequential pattern mining, TALENT \cite{chen2023talent} incorporates the concept of non-overlapping sequential patterns into target queries. Recently, Hu \textit{et al.} \cite{hu2024targeted} considered the continuity features in sequence data and proposed the TCSPM algorithm based on sequence segmentation operations and query sequence pruning strategies. The main contributions and characteristics of the existing targeted mining algorithms mentioned above are summarized in Table \ref{table: Comparison}.

Until now, numerous applications have shown a heightened sensitivity to timing accuracy, emphasizing the importance of identifying precisely located rules. Targeted mining enhances the specificity and effectiveness of the returned results. Therefore, in this paper, we propose the concept of precise localization with targeted queries and introduce a novel algorithm named TaMIPER.

\begin{table}[ht]
    \small
    \centering
    \footnotesize
    % \captionsetup{labelfont={color=red}}
    \caption{Comparison of related work.}
    \label{table: Comparison}
    \begin{tabularx}{0.5\textwidth}
    {|m{1.8cm}<{\raggedright}
    |m{6.37cm}<{\raggedright}|}
        \toprule
        \hline
         \textbf{Algorithm} & \multicolumn{1}{c|}{\textbf{Description}} \\
         \hline
         TargetUM \cite{miao2022targeted} & The first target-based high-utility itemset mining algorithm. It uses a lexicographic querying tree and three pruning strategies. \\ 
         \hline
         TaSPM \cite{huang2024taspm} & The first targeted SPM algorithm, which introduces several pruning strategies to enhance efficiency and reduce unnecessary operations in the mining process. \\ 
         \hline
         TaSRM \cite{gan2022towards} & The first targeted sequential rule mining algorithm. It introduces various pruning strategies and optimizations. \\ 
         \hline
         TUSQ \cite{zhang2022tusq} & The targeted high-utility sequence querying algorithm. It uses several novel upper bounds and a compact data structure. \\ 
         \hline
         TALENT \cite{chen2023talent} & An algorithm for targeted non-overlapping SPM. It utilizes breadth-first and depth-first searching methods, along with pruning strategies. \\ 
         \hline
         TCSPM \cite{hu2024targeted} & It addresses the problem of target-oriented contiguous SPM and utilizes the data structure named sequence chain and a reverse matching technique. \\ 
         \hline
         \bottomrule
        \end{tabularx}
\end{table}

\section{Preliminaries}  \label{sec: preliminaries}

In this section, we introduce several fundamental definitions and concepts related to the precise mining of target episode rules. Furthermore, we formulate the problem of targeted mining.

\begin{definition}[Event and event sequence \cite{ao2017mining}]
   \rm An event is defined as a specific activity or occurrence that takes place at a particular moment in time. For a finite set of events $\xi$, an event sequence is defined as $S$ = $<$($E_1$, $t_1$), ($E_2$, $t_2$), \dots, ($E_n$, $t_n$)$>$, where $E_i$ belongs to $\xi$, and $t_i$ represents the timestamp of the event $E_i$. Moreover, for any $i$ and $j$, where 1 $\leq$ $i$ $<$ $j$ $\leq$ $n$, it holds that $t_i$ $<$ $t_j$.
\end{definition}

For example, Fig. \ref{fig:events} illustrates an event sequence $S$ = $<$(\{$D$\}, 1), (\{$A$, $D$\}, 3), (\{$B$, $A$\}, 4), (\{$D$\}, 5), (\{$A$, $B$\}, 6), (\{$B$, $C$, $E$\}, 7), (\{$E$\}, 8), (\{$E$\}, 9), (\{$A$, $F$\}, 10), (\{$A$, $D$\}, 11), (\{$F$\}, 12)$>$.

\begin{definition}[Episode and episode occurrence \cite{ao2017mining}]
   \rm An episode $\alpha$ is defined as a non-empty ordered sequence in the form of $<$$e_{a1}$, \dots, $e_{ai}$, \dots, $e_{an}$$>$, where for all $i$ $\in$ [1, n], $e_{ai}$ belongs to $\xi$. Furthermore, within episode $\alpha$ for all 1 $\leq$ $i$ $<$ $j$ $\leq$ $n$, the event $e_{aj}$ occurs after event $e_{ai}$. An $n$-episode, where $n$ represents its length, is an episode of length $n$. The occurrence of episode $\alpha$ is defined as the time intervals of its occurrences, represented as [$t_{a1}$, \dots, $t_{ai}$, \dots, $t_{an}$], which can be abbreviated as [$t_{a1}$, $t_{an}$]. For all $i$ $\in$ [1, $n$], the event $e_{ai}$ occurs at time $t_{ai}$, and for all 1 $\leq$ $i$ $<$ $j$ $\leq$ $n$, $t_{ai}$ $<$ $t_{aj}$. Finally, the \textit{ocSet}($\alpha$) is defined as the set of all occurrences of episode $\alpha$ in the event sequence.
\end{definition}

For example, in Fig. \ref{fig:events}, \textit{ocSet}($<$$E$, $F$, $D$$>$) = \{[7, 10, 11], [8, 10, 11], [9, 10, 11]\}.

\begin{definition}[Minimal episode occurrence (MEO) \cite{ao2017mining}]
   \rm Minimal episode occurrence is defined as the minimal episode occurrence within the set of episode occurrences for the same episode with the same end time. If [$t_{ai}$, $t_{aj}$] is the minimal episode occurrence for episode $\alpha$, then there does not exist [$t_{ai}$', $t_{aj}$] as an episode occurrence for $\alpha$, where $t_{ai}$' $<$ $t_{aj}$. The \textit{\textit{moSet}}($\alpha$) is defined as the set of all occurrences of minimal episode $\alpha$ in the event sequence.
\end{definition}

For example, for \textit{moSet}($<$$E$, $F$, $D$$>$) = \{[7, 11], [8, 11], [9, 11]\}, both [7, 11] and [8, 11] encompass [9, 11], thus they are not MEO. Only [9, 11] qualifies as an MEO, denoted as \textit{moSet}($<$$E$, $F$, $D$$>$) = {[9, 11]}.

\begin{definition}[Fixed-gap episode and fixed-gap episode occurrence (FEO) \cite{ao2017mining}]
   \rm The form of a fixed-gap episode $\beta$ is defined as ($<$$e_{b1}$, \dots, $e_{bi}$, \dots, $e_{bn}$$>$, $<$$\Delta$$t_1$, \dots, $\Delta$$t_i$, \dots, $\Delta$$t_{n-1}$$>$) where, for any $i$ $\in$ [1, $n$ - 1], $e_{bi}$ belongs to $\xi$, and $\Delta$$t_i$ is the time span between $e_{bi}$ and $e_{bi+1}$. For all 1 $\leq$ $i$ $<$ $j$ $\leq$ $n$, $e_{bj}$ occurs after $e_{bi}$. A fixed-gap $n$-episode is defined as a fixed-gap episode of length $n$. For a fixed-gap episode $\beta$, [$t_{b1}$, \dots,$t_{bi}$, \dots, $t_{bn}$] represents its occurrences (FEO). Here, for all $i$ $\in$ [1, $n$ - 1], $t_{i+1}$ $>$ $t_{i}$, and $\Delta$$t_i$ = $t_{i+1}$ – $t_{i}$. 
\end{definition}

For example, in Fig. \ref{fig:events}, ($<$$E$, $F$$>$, $<$3$>$) is a fixed-gap episode where the events $E$ and $F$ have a time span of 3. The \textit{ocSet}($\beta$) is defined as the collection of all occurrences of episode $\beta$ on the event sequence. Therefore, \textit{ocSet}($<$$E$, $F$$>$,$<$3$>$) = \{[7, 10], [9, 12]\}.

\begin{definition}[Support of minimal episode and support of fixed-gap episode \cite{ao2017mining}]
    \label{definition: minSup}
    \rm The minimum episode support is defined as the number of distinct MEOs, denoted as \textit{sp}($\alpha$) = $\vert$\textit{moSet}($\alpha$)$\vert$. It is considered frequent when the minimum episode support is not less than the user-specified threshold. Similarly, the support of the fixed-gap episode is defined as the number of distinct FEOs, denoted as \textit{sp}($\beta$) = $\vert$\textit{ocSet}($\beta$)$\vert$. It is worth noting that episodes composed of the same events with different gaps are not considered the same fixed-gap episodes.
\end{definition}

For example, consider the minimum episode $<$$E$, $F$, $D$$>$, its support \textit{sp}($<$$E$, $F$, $D$$>$) = $\vert$moSet($<$$E$, $F$, $D$$>$)$\vert$ = 1. If the user specifies the threshold \textit{minSup} = 1, then $<$$E$, $F$, $D$$>$ is frequent. However, if \textit{minSup} = 2, then $<$$E$, $F$, $D$$>$ is not frequent. As for fixed-gap episodes, such as ($<$$E$, $F$, $D$$>$, $<$2, 1$>$) (occurring in the time interval [8, 11]) and ($<$$E$, $F$, $D$$>$, $<$1, 1$>$) (occurring in the time interval [9, 11]), these two episodes are distinct. However, for fixed-gap episodes like ($<$$D$, $A$, $B$$>$, $<$2, 1$>$) (occurring in the time interval [1, 4]) and ($<$$D$, A, $B$$>$, $<$2, 1$>$) (occurring in the time interval [3, 6]), they occur at different times but represent the same fixed-gap episode.

\begin{definition}[Super-episode and superset of episode \cite{huang2024taspm}]
    \rm Given an episode $\alpha$, if there exists another episode $\beta$ that satisfies the following conditions: (1) Episode $\beta$ contains all the events present in episode $\alpha$. (2) In episode $\beta$, the order of events that are the same as in episode $\alpha$ remains consistent. Then we refer to episode $\beta$ as one of the super-episodes of $\alpha$. Note that episode $\beta$ can be equal to or greater in length than episode $\alpha$. The key criteria are that $\beta$ includes all the events from $\alpha$, and the order of event occurrences in $\beta$ matches the order in which they occur in $\alpha$. In an event sequence, the set of all super-episodes of $\alpha$ is referred to as the super-sets of $\alpha$.
\end{definition}

For example, $<$$H$, \_, $G$$>$ is not a super-episode of $<$$G$, $H$$>$ because, even though $<$$H$, \_, $G$$>$ contains all the events from $<$$G$, $H$$>$, the order of events $G$ and $H$ in $<$$H$, \_, $G$$>$ is different from $<$$G$, $H$$>$. Assuming that the super-episodes of $<$$G$, $H$$>$ include $<$\_, $G$, $H$$>$, $<$$G$, $H$, \_$>$, $<$$G$, \_, $H$$>$, $<$$G$, \_, \_, $H$, \_$>$, and $<$$G$, $H$$>$, the superset of the episode $<$$G$, $H$$>$, denoted as \textit{superSet}($<$$G$, $H$$>$) = \{$<$\_, $G$, $H$$>$, $<$$G$, $H$, \_,$>$, $<$$G$, \_, $H$$>$, $<$$G$, \_, \_, $H$, \_$>$, $<$$G$, $H$$>$\}.

\begin{definition}[Episode rule \cite{ao2017mining}]
    \label{definition: deltaAndEpsilon}
    \rm Consider two episodes, $\alpha$ and $\beta$, represented as ($<$$e_{a1}$, \dots, $e_{ai}$, \dots, $e_{an}$$>$, [$t_{a1}$, \dots, $t_{ai}$, \dots, $t_{an}$]) and ($<$$e_{b1}$, \dots, $e_{bi}$, \dots, $e_{bn}$$>$, [$t_{b1}$, \dots, $t_{bi}$, \dots, $t_{bn}$]) respectively. When defining the episode rule $\alpha$ $\rightarrow$ $\beta$, the following conditions must be met: (1) The occurrence time of the first event($e_{b1}$) in episode $\beta$  must be later than the occurrence time of the last event($e_{an}$) in episode $\alpha$; (2) $\alpha$, as the antecedent of the rule, needs to satisfy the time span requirement, that is, $t_{an}$ - $t_{a1}$ $<$ $\delta$; (3) $\beta$, as the consequent of the rule, must adhere to the time span restriction, that is, $t_{bn}$ - $t_{an}$ $\leq$ $\epsilon$. It is important to emphasize that $t_{an}$ represents the occurrence time of the last event in episode $\alpha$.
\end{definition}

For example, assuming $\delta$ = 2 and $\epsilon$ = 4, in Fig. \ref{fig:events}, the episode ($<$$D$, $B$$>$[5, 6]) $\rightarrow$ ($<$$C$, $F$$>$[7, 10]) meets the criteria, whereas ($<$$D$, $B$$>$[3, 6]) $\rightarrow$ ($<$$C$, $F$$>$[7, 10]) and ($<$$D$, $B$$>$[3, 4]) $\rightarrow$ ($<$$C$, $F$$>$[7, 10]) do not. This is because for ($<$$D$, $B$$>$[3, 6]), there is $\delta$ = 6 - 3 = 3, which is greater than 2 and doesn't meet the requirement for the antecedent time span ($\delta$). Similarly, for ($<$$D$, $B$$>$[3, 4]) $\rightarrow$ ($<$$C$, $F$$>$[7, 10]), there is $\epsilon$ = 10 - 4 = 6, which is greater than 4 and doesn't comply with the consequent time span ($\epsilon$).

\begin{definition}[Target episode and target episode mining]
    \label{definition: Qe}
    \rm Consider a query episode $\mathcal{Q}_e$. The target episode $\mathcal{T}_e$ must satisfy the condition that $\mathcal{Q}_e$ is a sub-episode of $\mathcal{T}_e$, meaning $\mathcal{Q}_e$ is contained within $\mathcal{T}_e$. The primary goal of target episode mining is to discover episode rules of the form $\alpha$ $\rightarrow$ $\beta$, where episode $\beta$ represents the target episode $\mathcal{T}_e$. Note that if only episode $\alpha$ contains $\mathcal{Q}_e$ and episode $\beta$ does not, then this episode rule is not the target rule we are seeking.
\end{definition}

For example, in Fig. \ref{fig:events}, without considering episode time spans and confidence conditions, assume that $\mathcal{Q}_e$ = $<$$C$, $F$$>$. Then, we would identify target episodes such as $<$$C$, $E$, $F$$>$, $<$$C$, $E$, $E$, $F$$>$, and more. Among these, one of the target episode rules would be $<$$A$, $D$, $A$$>$ $\rightarrow$ $<$$C$, $E$, $F$$>$.

\begin{definition}[Targeted precise-positioning episode rule]
    \rm Targeted precise-positioning episode rule (TaPER) is based on the concept of episode rule, formulated as $\alpha$ $\overset{\Delta t}{\rightarrow}$ $\beta$, where episode $\alpha$ = $<$$e_{a1}$, \dots, $e_{ai}$, \dots, $e_{an}$$>$ represents the minimal episode that frequently occurs, and episode $\beta$ = ($<$$e_{b1}$, \dots, $e_{bi}$, \dots, $e_{bn}$$>$, $<$$t_{b1}$, \dots, $t_{bi}$, \dots, $t_{bn}$$>$) represents a fixed-gap episode that must include the query episode. In other words, $\beta$ is the target episode. $\Delta$t is the time interval that has elapsed for episode $\alpha$ before episode $\beta$ occurs. Both episodes $\alpha$ and $\beta$ must adhere to the previously mentioned constraints on episode time spans, $\delta$ and $\epsilon$.
\end{definition}

For example, in Fig. \ref{fig:events}, without considering confidence conditions, assume that $\mathcal{Q}_e$ = $<$$C$, $F$$>$, $\delta$ = 4, and $\epsilon$ = 4. In this case, $<$$A$, $D$, $A$$>$ $\overset{1}{\rightarrow}$($<$$C$, $E$, $F$$>$,$<$1,2$>$) represents a valid TaPER. This rule signifies that if the episode $<$$A$, $D$, $A$$>$ occurs, then after the time interval of 1, the episode $<$$C$, $E$, $F$$>$ occurs, containing the desired $\mathcal{Q}_e$. In the episode $<$$C$, $E$, $F$$>$, the event $C$ occurs the time interval of 1 later than the event $E$; two time intervals later, the event $F$ occurs.

\begin{definition}[Occurrence of TaPER \cite{ao2017mining}]
    \rm For the TaPER $\Gamma$ = $\alpha$ $\overset{\Delta t}{\rightarrow}$ $\beta$, where the MEO is ($\alpha$, [$t_{a1}$, $t_{ak}$]), and the FEO is ($\beta$, ($t_{b1}$, $t_{bn}$)), we define [$t_{a1}$, $t_{ak}$, $t_{b1}$, $t_{bn}$] as the occurrences of $\Gamma$ and \textit{OcSet}($\Gamma$) as the set of all occurrences of $\Gamma$.
\end{definition}

\begin{definition}[Support and confidence of TaPER \cite{ao2017mining}]
    \label{definition: confidence}
    \rm The support of the TaPER $\Gamma$ is defined as the number of distinct occurrences of $\Gamma$, denoted by \textit{sp}($\Gamma$) = $\vert$\textit{OcSet}($\Gamma$)$\vert$. $\Gamma$'s confidence is defined as \textit{minConf}($\Gamma$) = $\frac{\text{sp}(T)}{\text{sp}(\alpha)}$. A TaPER is considered valid if and only if $\alpha$ is frequent and \textit{minConf}($\Gamma$) is not less than the user-specified minimum confidence.
\end{definition}

For example, in Fig. \ref{fig:events}, assuming $\delta$ = 2 and $\epsilon$ = 4, the support of $\Gamma$ = ($<$$A$, $B$$>$ $\overset{1}{\rightarrow}$ ($<$$E$, $F$, $D$$>$, $<$2, 1$>$)) is 1. Its occurrences are [6, 7, 8, 10, 11], respectively. For $\Gamma$ = ($<$$A$, $B$$>$ $\overset{1}{\rightarrow}$ ($<$$E$, $F$, $D$$>$, $<$2, 1$>$)), given that \textit{sp}($\alpha$) = 3 and \textit{sp}($\Gamma$) = 1, the \textit{minConf}($\Gamma$) is 33.3\%.

\textbf{Problem statement:} Given an event-based sequence, the problem of targeted mining of precise-positioning episode rules is to discover all valid TaPERs, where each rule must satisfy user-specified parameters:  $\mathcal{Q}_e$ (Definition \ref{definition: Qe}): Each rule must include the target episode; $\delta$ (Definition \ref{definition: deltaAndEpsilon}): The span of the antecedent is less than $\delta$; $\epsilon$ (Definition \ref{definition: deltaAndEpsilon}): The span of the consequent is no greater than $\epsilon$; \textit{minSup} (Definition \ref{definition: minSup}): The support of the antecedent is not less than \textit{minSup}; and \textit{minConf} (Definition \ref{definition: confidence}): The rule's confidence is not lower than \textit{minConf}.

\section{Proposed Algorithm} \label{sec: algorithm}
In this section, we introduce TaMIPER, an efficient tree-based algorithm designed to enhance the storage of TaPER, expedite the mining of rules, and address the problem of targeted mining of precise-positioning episode rules.

\subsection{TaMIPER Framework}

Drawing inspiration from prior research \cite{ao2017mining, huang2024taspm}, TaMIPER is grounded in a tree-based data structure and comprises two pivotal phases. The initial phase involves mining the superset of $\mathcal{Q}_e$, which provides essential positional information for potential target rules. In the following phase, it first extracts frequent MEOs, using them as antecedents to efficiently derive TaPER. Additionally, the positional information obtained during pre-expansion is employed for pruning purposes.

In Algorithm \ref{alg: TaMIPER}, the TaMIPER algorithm's mining process is delineated. It initiates by invoking the GetSuperSet function to identify all super-episodes of $\mathcal{Q}_e$ and their positions in the entire event sequence. This function takes three inputs: the event sequence $S$, the length threshold $\delta$ for an antecedent, and the query episode $\mathcal{Q}_e$ (line 1). Following this, the set of frequent MEOs is extracted across the entire event sequence (line 2). Each MEO serves as an antecedent episode for potential rules, and the MiningFEO procedure is called to extract effective TaPERs (lines 3-7). The inputs of MiningFEO include the event sequence $S$, the query episode $\mathcal{Q}_e$, a potential antecedent $\alpha$, the end-time set of antecedent MEO $\mathcal{ET}_a$, the superset of $\mathcal{Q}_e$ \textit{superSet}, the minimum frequency threshold for effective TaPERs \textit{minConf}, and the length threshold for subsequent events $\epsilon$.

\begin{algorithm}[ht] 
    \small
    \caption{The TaMIPER algorithm}
    \label{alg: TaMIPER}
    \LinesNumbered
    \KwIn{$S$: the event sequence; \textit{minSup}: the minimum support threshold; \textit{minConf}: the minimum confidence threshold; $\delta$: the length of the antecedent episode; $\epsilon$: the length of the consequent episode; and $\mathcal{Q}_e$: the query episodes for consequence.}
    \KwOut{$R$: the set of valid TaPER.}
    
    \textit{superSet} $\leftarrow$ \textbf{call} \textbf{GetSuperSet}($S$, $\delta$, $\mathcal{Q}_e$)\;
    $A$ $\leftarrow$ extract frequent minimal-occurrence episodes from $S$ with the specified thresholds of $\delta$ and \textit{minSup}\;

    \For{\rm each $\alpha$ in $A$}{
        $\mathcal{ET}_a$ $\leftarrow$ the collection of end time of each MEO of $\alpha$\;
        $R$ $\leftarrow$ $\varnothing$\;
        $R$ $\leftarrow$ $R$ $\cup$ \textbf{call} \textbf{MiningFEO}($\mathcal{S}$, $\mathcal{Q}_e$, $\alpha$, $\mathcal{ET}_a$, \textit{superSet}, $\textit{sp}(\alpha) \times \textit{minConf}$, $\epsilon$)\;
    }
    \textbf{return} $R$
\end{algorithm}

\subsection{Mining the superset of $\mathcal{Q}_e$}

\begin{property}[Occurrence interval property of the antecedent] 
\label{property: antecedent}
    Consider an episode $\beta$($<$$e_{b1}$, \dots, $e_{bn}$$>$, [$t_{b1}$, \dots, $t_{bn}$]) and a time span threshold $\epsilon$. If a TaPER's subsequent episode includes episode $\beta$, then the possible occurrence interval for the antecedent of that TaPER is [$t_{bn}$ - $\epsilon$, $t_{b1}$), where $t_{bn}$ – $\epsilon$ is replaced by 1 when $t_{bn}$ – $\epsilon$ is less than 1
\end{property}

\begin{proof}
    For each rule, such as ($<$$e_{a1}$, \dots, $e_{an}$$>$, [$t_{a1}$, \dots, $t_{an}$]) $\overset{\Delta t}{\rightarrow}$ ($<$$e_{b1}$, \dots, $e_{bn}$$>$, [$t_{b1}$, \dots, $t_{bn}$]), it must satisfy $t_{bn}$ – $t_{an}$ $\leq$ $\epsilon$. Therefore, given an episode $\beta$($<$$e_{b1}$, \dots, $e_{bn}$$>$, [$t_{b1}$, \dots, $t_{bn}$]) and a time span threshold $\epsilon$, the antecedent for episode $\beta$ should satisfy $t_{an}$ $\geq$ $t_{bn}$ – $\epsilon$. Additionally, since there is a time gap of $\Delta t$ between $e_{an}$ and $e_{b1}$, and $\Delta t$ is greater than or equal to 1, $t_{an}$ must be less than $t_{b1}$. In summary, $t_{an}$ must satisfy the ending time occurrence interval for episode $\beta$ as [$t_{bn}$ – $\epsilon$, $t_{b1}$). Since moments are positive integers, $t_{bn}$ – $\epsilon$ is replaced by 1 when $t_{bn}$ – $\epsilon$ is less than 1.
\end{proof}

In Algorithm \ref{alg: GetSuperSet}, the process of mining the superset of $\mathcal{Q}_e$ in the input sequence is introduced. The algorithm iterates through each event occurrence at every timestamp, from the beginning to the end of the input event sequence, to obtain an event set $E$ (lines 1-2). Next, the events in $E$ that are not present in $\mathcal{Q}_e$ are removed (line 3). Subsequently, the event set $E$ and its corresponding \textit{timestamps} are added to the combination \textit{graphList} (line 4). After inclusion in the \textit{graphList}, its internal nodes need to be combined into a super-episode of $\mathcal{Q}_e$ (lines 5-23). If the number of nodes in the \textit{graphList} is greater than 2, a backward traversal of the \textit{graphList} is performed, except for the last newly added node. Specifically, the backward traversal starts at the second-to-last node and ends at the first node (lines 5-6). The \textit{prefixSets} are derived from each node, and for each \textit{prefix}, the \textit{episodeSet} that have already been combined are retrieved (lines 7-9). Each event $e$ from the event set $E$ is then added to each \textit{episode} (lines 10-12). If the \textit{newEpisode} length equals the length of $\mathcal{Q}_e$, the prefix event \textit{prefix}, the event $e$, and the occurrence times of the two events are stored in the superset of $\mathcal{Q}_e$, \textit{superSet}. It is worth noting that the data structure for \textit{superSet} is a key-value pair, where the key is a string representing the event, and the value corresponds to a set of multiple integer lists used to record timestamps. In this case, \textit{prefix} corresponds to the first event in the \textit{newEpisode}, and its timestamp is the occurrence time of the first event. The parameter $e$ represents the last event in the \textit{newEpisode}, and its timestamp corresponds to the occurrence time of the last event. Otherwise, it is added to the \textit{episodeSet} for that \textit{prefix} (lines 13-18). During the traversal at each timestamp, nodes outside the time range are removed from the \textit{graphList} based on $\epsilon$ (line 24). This is because $\epsilon$ is the temporal constraint threshold for the subsequent. If the superset of $\mathcal{Q}_e$ does not meet the $\epsilon$ condition, it fails to satisfy the criteria of the targeted precise-positioning episode rules.

The subsequent steps involve obtaining all occurrence time intervals for the superset of $\mathcal{Q}_e$ and calculating the possible occurrence range of the antecedent for each episode according to Property \ref{property: antecedent} (lines 26-33). First, $\mathcal{Q}_e$ is used as a key to retrieve all occurrence intervals, and then, based on the occurrence time of the last event in $\mathcal{Q}_e$, \textit{time[1]}, and the subsequent length threshold, $\epsilon$, the possible occurrence time range of the antecedent is calculated as [\textit{time[0]}, \textit{range}) and stored in \textit{time}. Finally, the superset of $\mathcal{Q}_e$ along with all their occurrence intervals are returned.

\begin{algorithm}[ht]
    \small
    \caption{GetSuperSet}
    \label{alg: GetSuperSet}
    \LinesNumbered
    \KwIn{$S$: the event sequence;  $\mathcal{Q}_e$: the query episodes for consequence; $\epsilon$: the length of the consequent episode.}
    \KwOut{\textit{superSet}: the superset of $\mathcal{Q}_e$ in $S$.}

    \For{\rm \textit{timeStamp} = $S.beginTime$ to $S.endTime$}{
        scan the corresponding \textit{timeStamp} of $S$ and get an event set $E$\;
        delete events in $E$ that do not exist in $\mathcal{Q}_e$\;
        add $E$ and \textit{timeStamp} to \textit{graphList}\;
        \If{\rm \textit{SizeOf}(\textit{graphList}) $>$ 2}{
            \For{\rm \textit{i} = \textit{SizeOf}(\textit{graphList}) - 2 down to 0}{
                \textit{prxfixSet} $\leftarrow$ \textit{graphList[i]}\;
                 \For{\rm each \textit{prefix} in \textit{prxfixSet}}{
                    \textit{episodeSet} $\leftarrow$ \textit{GetEpisodeSet}(\textit{prefix})\;
                    \For{\rm each $e$ in $E$}{
                        \For{\rm each \textit{episode} in \textit{episodeSet}}{
                            \textit{newEpisode} $\leftarrow$ \textit{episode} + $e$\;
                            \If{\rm \textit{Length}(\textit{newEpisode}) $>$ \textit{SizeOf}(\textit{$\mathcal{Q}_e$})}{
                                save $e$, \textit{timeStamp}, \textit{prefix} and the timestamp corresponding to \textit{prefix} in \textit{superSet}\;
                            }
                            \Else{
                                \textit{AddTo}(EpisodeSet(\textit{prefix}), \textit{newEpisode})\;
                            }
                        }
                    }
                 }
            }
        }
        delete the nodes in \textit{graphList} that exceed the threshold range based on $\epsilon$\;
    }
    \textit{timeList} $\leftarrow$ Get the value corresponding to key $\mathcal{Q}_e$ in \textit{superSet}\;
        \For{\rm each \textit{time} in \textit{timeList}}{
            \textit{range} $\leftarrow$ \textit{time[1]} - $\epsilon$\;
            \If{\rm \textit{range} $<$ 0}{
                \textit{range} = 1\;
            }
            \textit{Add}(\textit{time}, \textit{range})\;
        }
    \textbf{return} \textit{superSet}
\end{algorithm}

\subsection{Mining the TaPER-tree of the Antecedent $\alpha$}

\subsubsection{Structure of TaPER-tree}

The root node of the TaPER-tree represents a frequent minimal occurrence episode and is comprised of the fields \textit{r.episode} and the set of end times \textit{r.tlist}. The TaPER-tree, with episode $\alpha$ as the root node, is denoted as $\mathcal{T}_a$, where \textit{r.episode} stores the minimal episode occurrence $\alpha$, and \textit{r.tlist} records all its end times, defined as \textit{r.tlist} = {$t_j$ $\vert$ [$t_i$, $t_j$] $\in$ \textit{moSet}($\alpha$)}. Non-root nodes of the TaPER-tree consist of episodes and their sets of end times. For a non-root node $q$, represented as (\textit{q.episode}, \textit{q.tlist}), \textit{q.episode} captures the occurring episodes, and \textit{q.tlist} records the set of end times for episode $q$.

\begin{figure}[ht]
    \centering
    \includegraphics[scale=0.7]{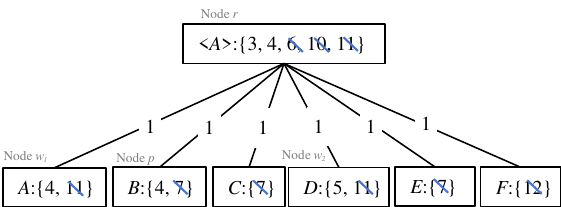}
    \caption{The TaPER-tree $\mathcal{T}_{<A>}$ when $i$ = 1 established by Algorithm \ref{alg: MiningFEO}.}
    \label{fig:T_A_1}
\end{figure}

Fig. \ref{fig:T_A_1} illustrates a TaPER-tree $\mathcal{T}_{<A>}$, with $<$A$>$ serving as the root node. The \textit{r.tlist} = \{$t_j$ $\vert$ [$t_i$, $t_j$] $\in$ \textit{moSet}($<$A$>$)\} = \{3, 4, 6, 10, 11\} represents the set of end times for $<$A$>$. Similarly, for non-root node $p$, \textit{p.event} records event $B$, and \textit{p.tlist} records the set of end times for event $B$.

\begin{figure}[ht]
    \centering
    \includegraphics[clip,scale=0.42]{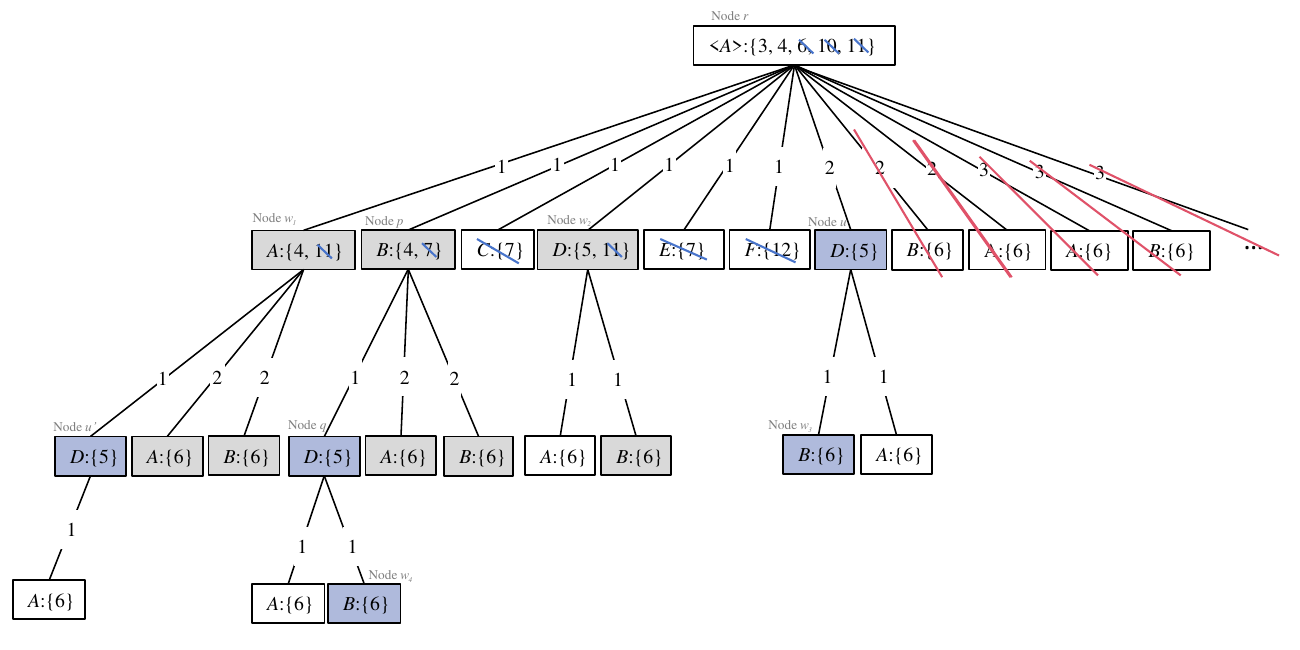}
    \caption{The TaPER-tree $\mathcal{T}_{<A>}$ when $i$ = 3 established by Algorithm \ref{alg: MiningFEO}.}
    \label{fig:T_A_5}
\end{figure}

The distance field on an edge between a parent node $p$ and a child node $q$ is defined as $d(p, q)$, representing the temporal separation between nodes $p$ and $q$. For any given $t_q$ $\in$ \textit{q.tlist}, there always exists a $t_p$ $\in$ \textit{p.tlist}, such that $t_q$ - $tp$ = $d(p, q)$. In Fig. \ref{fig:T_A_5}, consider the node $p$. Its distance to the root node $r$ is 1, meaning \textit{d(p, q)} = 1. Given \textit{r.tlist} = \{3, 4, 6, 10, 11\} and let \textit{tlist} represent the occurrence times after $r$. Therefore, it can be expressed as \textit{tlist} = \{4, 5, 7, 11, 12\}. As the event $B$ occurs in \textit{tlist} only at timestamps 4 and 7, for \textit{p.event} = $B$, \textit{p.tlist} = \{4, 7\}. Conversely, since \textit{p.tlist} = \{4, 7\} and \textit{d(r, p)} = 1, there must be \{3, 6\} $\subseteq$ \textit{r.tlist}.

For any TaPER-tree, the path from the root node to a non-root node can represent a TaPER. In other words, for a non-root node $q$, if the path from the root node $r$ to the node $q$ is $r$ $\rightarrow$ \textit{q1} $\rightarrow$ \textit{q2} $\rightarrow$ \dots $\rightarrow$ \textit{qn} $\rightarrow$ $q$, then the TaPER represented by node $q$ is r.episode $\overset{d(r, q1)}{\rightarrow}$ ($<$\textit{q1.event}, \textit{q2.event}, \dots, \textit{qn.event}, \textit{q.event}$>$, $<$\textit{d}(\textit{q1}, \textit{q2}), \dots, \textit{d}(\textit{qn}, $q$)$>$). In Fig. \ref{fig:T_A_5}, the path from the root node $r$ to the node $q$ represents the TaPER as $A$ $\overset{1}{\rightarrow}$ ($<$$B$, $D$$>$, $<$1$>$).

\begin{property}[Support counting \cite{ao2017mining}] 
\label{property: support}
    For every non-root node $q$ on the TaPER-tree, the support of each TaPER related to $q$ is equal to the cardinality of $q$ node's event end-time set, denoted as $\vert$\textit{q.tlist}$\vert$.
\end{property}
\begin{proof}
    Without loss of generality, we use nodes $p$ (parent) and $q$ (child) to represent any pair of parent-child nodes on the TaPER-tree. We use $r$ to denote the root node of the TaPER tree, and $\Gamma$ to denote the TaPER associated with node $q$.

   (i) For any $j$ in \textit{q.tlist}, there exists $i$ in \textit{p.tlist} such that $i$ + \textit{d(p, q)} = $j$. In other words, for each distinct \textit{q.event} timestamp in \textit{q.tlist}, there is a corresponding occurrence of \textit{p.event} in \textit{p.tlist} that precedes it by $d(p, q)$ time units. Therefore, by backtracking, we can identify complete rules for each timestamp in \textit{q.tlist}, implying that \textit{sp}($\Gamma$) $\geq$ $\vert$\textit{q.tlist}$\vert$.

   (ii) Assuming there are other $\Gamma$, that are not included in \textit{q.tlist}, we can then identify additional end-times for \textit{r.episode} not present in \textit{r.tlist}. However, based on the definition of \textit{r.tlist}, \textit{r.tlist} = \{$t_j$ $\vert$ [$t_i$, $t_j$] in \textit{moSet}(\textit{r.episode})\}. This implies that \textit{r.tlist} contains all the end-times, contradicting the assumption. Therefore, the assumption is invalid, and we conclude that \textit{sp}($\Gamma$) $\leq$ $\vert$\textit{q.tlist}$\vert$. In summary, we have \textit{sp}($\Gamma$) = $\vert$\textit{q.tlist}$\vert$.
\end{proof}

The process of mining the TaPER-tree $\mathcal{T}_a$ regarding the episode $\alpha$ is, in fact, the process of mining precise-positioning rules with episode $\alpha$ as the antecedent. This process commences with episode $\alpha$ as the root node and begins mining subsequent events. In Fig. \ref{fig:T_A_1}, where $i$ = 1 represents a search distance of 1, it entails searching for all events occurring $i$ units of time after the end time of episode $\alpha$. If the events meet the threshold set by the user, they are added to $\mathcal{T}_a$, signifying the generation of a new rule. The value of $i$ is incremented one by one, and the search stops when $i$ $>$ $\epsilon$. In Fig. \ref{fig:T_A_1}, since nodes $p$ and $w$ satisfy the relevant threshold conditions, they are added to $\mathcal{T}_a$. This means that new rules have been generated: $<$A$>$$\overset{1}{\rightarrow}$$<$B$>$ and $<$A$>$$\overset{1}{\rightarrow}$$<$$D$$>$. 

\subsubsection{Pruning strategy of TaPER-tree}

\begin{strategy}[Pre-expansion-Based Pruning Strategy (PBPS)]      \label{strategy: PBPS} 
    \rm For each moment within the end-time set of the antecedent that does not fall within any of the potential pre-expansion time intervals, a filtering process is initiated. By scanning the entire input sequence, we extract the super-episode of $\mathcal{Q}_e$ within the event sequence, along with their respective occurrence times. Specifically, this refers to the occurrence times of the first and last events in the superset of $\mathcal{Q}_e$. This enables the deduction of the time intervals that each antecedent event must satisfy according to Property \ref{property: antecedent}. If the end-time of an antecedent fails to meet all possible occurrence intervals, it indicates that beyond that moment, effective TaPERs containing $\mathcal{Q}_e$ cannot be mined. Therefore, we proactively implement filtering at that timestamp.
\end{strategy}

Assume $\mathcal{Q}_e$ = $<$$D$, $A$$>$, $\delta$ = 2, $\epsilon$ = 4, \textit{minSup} = 2, \textit{minConf} = 0.01, and use the input sequence from the event sequence in Fig. \ref{fig:events}. Since \textit{sp}($<$A$>$) = 5 $>$ \textit{minSup}, the episode $<$A$>$ is frequent and can be used as an antecedent to generate rules, as shown in Fig. \ref{fig:T_A_1}. Specifically, $<$A$>$ serves as the root node $r$, and \textit{r.tlist} represents the end-time set of $<$A$>$. However, the super-episode of $\mathcal{Q}_e$ only appears in [1, 3], [1, 4], [3, 4], [3, 6], and [5, 6]. According to Property \ref{property: antecedent}, the pre-expansion time intervals are [1, 3), [2, 3), and [2, 5). Consequently, the end times in \textit{r.tlist} that do not satisfy all possible intervals can be proactively filtered out. In other words, the end times 6, 10, and 11 are filtered out, leaving end times 3 and 4 for further rule mining.

\begin{strategy}[Distance-Based Pruning Strategy (DBPS)]      \label{strategy: DBPS} 
    \rm During the process of mining effective rules from the antecedent, nodes with search distances exceeding the maximum distance are precluded from being incorporated into the TaPER-tree for the expansion of their descendant nodes and subsequent rule generation. For each occurrence timestamp within the antecedent that adheres to the pre-expansion time, the distance between the end-time of the antecedent and the start-time of the first event in the super-episode of $\mathcal{Q}_e$ is calculated. This distance represents the antecedent's distance from the super-episode. After determining the search distances for all antecedent timestamps, the maximum search distance is computed. If, during the scanning process, the search distance surpasses the maximum search distance, it signifies that in subsequent mining endeavors, the super-episode of $\mathcal{Q}_e$ will no longer be uncovered. In other words, the nodes obtained in subsequent scans will not evolve into effective rules containing $\mathcal{Q}_e$. Consequently, there is no necessity to include them in the TaPER-tree for rule generation.
\end{strategy}

Similarly, we can deduce the pre-expansion time intervals [1, 3), [2, 3), and [2, 5). For the end times 3 and 4, as they conform to the pre-expansion interval [2, 5), the maximum search distance is calculated as 5 - 3 = 2. Consequently, when the search distance equals or surpasses 2, nodes with search events other than the first event $D$ of $\mathcal{Q}_e$ are excluded from being added to the TaPER-tree for rule generation. As illustrated in Fig. \ref{fig:T_A_5}, except for the node $u$, subsequent nodes are not incorporated into the TaPER-tree due to their excessive search distance.

\begin{strategy}[Node-Based Pruning Strategy (NBPS)]      \label{strategy: NBPS} 
    \rm For nodes that cannot grow the super-episode of $\mathcal{Q}_e$, they are deactivated. It is understood that each $\mathcal{Q}_e$ comprises a start event and an end event. When the first event of the last episode in the superset of $\mathcal{Q}_e$ enters the TaPER-tree, the existing nodes in the TaPER-tree will no longer expand to generate new first events for episodes that include $\mathcal{Q}_e$. Therefore, we only maintain the activity of nodes in the TaPER-tree related to the first event of $\mathcal{Q}_e$ and their descendant nodes. All other nodes are deactivated, preventing them from generating descendant nodes and rules. Similarly, when the last event of the last episode in the superset of $\mathcal{Q}_e$ enters the TaPER-tree, all events related to the super-episode of $\mathcal{Q}_e$ will no longer appear. Consequently, we can halt the generation of rules that do not include $\mathcal{Q}_e$, meaning we stop nodes (nodes along the path from the root node to that node that do not contain $\mathcal{Q}_e$) from expanding their descendants. 
\end{strategy}

Assuming the same conditions as before, for the end times 3 and 4 that satisfy the pre-expansion interval [2, 5), the maximum search distance is 5 - 3 = 2. In other words, when the search distance is 2, after the last super-episode of $\mathcal{Q}_e$, with the first event $D$, enters the TaPER-tree, nodes that do not contain the first event $D$ are deactivated and no longer generate descendant nodes. Specifically, in Fig. \ref{fig:T_A_5}, all nodes except $q$, $u$, $u$', $w_2$, and their descendant nodes are deactivated. Similarly, when the last super-episode of $\mathcal{Q}_e$, with the final event $A$, enters the TaPER-tree, if a node's path (from the root node to that node) does not contain event $D$ and event $A$, such as nodes $u$, $u$', $q$, $w_3$, and $w_4$ in Fig. \ref{fig:T_A_5}, that node is deactivated, and no more descendant nodes are generated.

\begin{property}[Length property of TaPER] 
\label{property: length}
    Given the query episode $\mathcal{Q}_e$, for all the valid TaPERs regarding the antecedent $\alpha$, they must all have \textit{Length}(TaPER) $\geq$ \textit{Length}($\alpha$) + \textit{Length}($\mathcal{Q}_e$).
\end{property}
\begin{proof}
    For all valid TaPERs related to the antecedent $\alpha$, as each TaPER must contain at least the antecedent $\alpha$ and the query episode $\mathcal{Q}_e$, we can conclude that the length of a TaPER must be greater than or equal to the sum of the lengths of the antecedent $\alpha$ and the query episode $\mathcal{Q}_e$, i.e., \textit{Length(TaPER)} $\geq$ \textit{Length}($\alpha$) + \textit{Length}($\mathcal{Q}_e$).
\end{proof}

\begin{strategy}[Length-Based Pruning Strategy (LBPS)] \label{strategy: LBPS}
    \rm Rules with a length less than the anticipated minimum target rule length are excluded from consideration. After implementing the three aforementioned pruning strategies, the TaPER-tree predominantly consists of rules containing $\mathcal{Q}_e$. However, in line with the characteristics of TaPER-tree, there exists a certain 'byproduct' during the process of generating target rules. Since each addition of a new node results in a new path, and thus the emergence of new rules, there are other rules generated with lengths shorter than the target rules prior to the adoption of target rules. As a result, we can filter out those rules with shorter lengths than the expected minimum rule length.
\end{strategy}

In Fig. \ref{fig:T_A_1}, with the same conditions, we can observe that even though the paths for each node (from the root node to that node) do not contain the complete $\mathcal{Q}_e$, they have generated invalid TaPERs. Therefore, we can perform effective pruning based on Strategy \ref{strategy: LBPS}. In other words, the rules represented by nodes such as $p$ and $w$, which have shorter lengths than the expected minimum rule length, are not considered for adoption.

\begin{algorithm}
    \small
    \caption{MiningFEO}
    \label{alg: MiningFEO}
    \LinesNumbered
    \KwIn{$S$: an event sequence; $\mathcal{Q}_e$: the query episodes for consequence; $\alpha$: a frequent minimal-occurrence episode; $\mathcal{ET}_a$: a set of end time of each MEO of $\alpha$; \textit{superSet}: the superset of $\mathcal{Q}_e$ in $S$; \textit{minFreq}: the minimum frequency threshold generating a valid TaPER; $\epsilon$: the length of the consequent episode.}
    \KwOut{$\mathcal{T}_a$: the complete TaPER-tree of antecedent $\alpha$.}

    % \textbf{(strategy \ref{strategy: PBPS})}\\
    \For{\rm each \textit{t} in $\mathcal{T}_a$}{
        \textit{presence} $\leftarrow$ using binary search to determine whether \textit{t} is present in \textit{superSet}\;
        \If{\rm \textit{presence} == true}{          
            add the distance from time \textit{t} to the first and last event in $\mathcal{Q}_e$ to \textit{spanFirst} and \textit{spanLast}\;
        }
        \Else{
            remove it from $\mathcal{ET}_a$\;
        }
    }
    \textit{maxFirst} $\leftarrow$ get the maximum value of \textit{spanFirst}\;
    \textit{maxLast} $\leftarrow$ get the maximum value of \textit{spanLast}\;
     construct the root node $r$ in $\mathcal{T}_a$ with \textit{r.episode} set to $\alpha$ and \textit{r.tlist} set to $\mathcal{ET}_a$\;
    \For{\rm \textit{i} = 1 to $\epsilon$}{
    % \textbf{(strategy \ref{strategy: NBPS})}\\
        \If{\rm \textit{i} == \textit{maxFirst} + 1}{          
            deactivate other nodes except for the $\mathcal{Q}_e$ first event and its descendant nodes\;
       }
        \ElseIf{\rm \textit{i} == \textit{maxLast} + 1}{
            deactivate other nodes except for the $\mathcal{Q}_e$ last event and its descendant nodes, \;
       }
       \textit{timeList} $\leftarrow$ $\{$\textit{t'} $\vert$ \textit{t} + \textit{i}, \textit{t} $\in$ \textit{r.tlist}$\}$\;
       examine each time stamp in \textit{timeList} to extract an event set \textit{Event'} where the frequency of each event in \textit{Event'} is not less than \textit{minFreq}\;
       \For{\rm \textit{event'} $\in$ \textit{Event'}}{
            \textit{u.event} $\leftarrow$ \textit{event'}\;
            \textit{u.tlist} $\leftarrow$ the time of occurrence of \textit{event'} within \textit{timeList}\;
            % \textbf{(strategy \ref{strategy: DBPS})}\\
            \If{\rm \textit{i} $<$ \textit{maxFirst}}{          
                add $u$ as a child node of $r$\;
            }
            \For{\rm each non-root node $w$ of $\mathcal{T}_a$ except $u$}{
                 \If{\rm \textit{IsDead}($w$)}{
                    continue\;
                 }
                 calculate \textit{d} as the distance between $r$ and $w$\;
                 \If{\rm \textit{i} $>$ \textit{d}}{
                    \textit{tmpList} = $\{$\textit{t'} $\vert$ \textit{t} + \textit{i} - \textit{d}, \textit{t} $\in$ \textit{w.tlist}$\}$\;
                    calculate $F$ as \textit{tmpList} $\cap$ \textit{u.tlist}\;
                    \If{\rm $F$ $>$ \textit{minFreq}}{
                        \textit{u'.event} $\leftarrow$ \textit{u.event}\; 
                        \textit{u'.tlist} $\leftarrow$ $F$\;
                        add $u$' as a child node of $w$\;
                        % \textbf{(strategy \ref{strategy: LBPS})}\\
                        decide whether to generate a new rule based on its length\;          
                    }
                 }
            }
       }
    }
    \textbf{return} $\mathcal{T}_a$
\end{algorithm}

In Algorithm \ref{alg: MiningFEO}, the complete process of mining the TaPER-tree for the antecedent episode $\alpha$ is introduced. The algorithm begins by employing Pruning Strategy \ref{strategy: PBPS} (lines 2-10). Based on the binary search, the algorithm determines whether each end time \textit{t} of the episode $\alpha$ exists within the pre-expansion interval. If it does, the distances between \textit{t} and the first and last events of $\mathcal{Q}_e$ are calculated and recorded as \textit{spanFirst} and \textit{spanLast}, respectively. Otherwise, the time \textit{t} is filtered out. Subsequently, the maximum search distances, \textit{maxFirst} and \textit{maxLast}, for the first and last events are obtained based on the search distances \textit{spanFirst} and \textit{spanLast} (lines 11-12). Next, a TaPER-tree $\mathcal{T}_a$ related to the episode $\alpha$ is constructed, with episode $\alpha$ as the root node (line 13). The algorithm then incrementally explores subsequent events under the condition of the time-span threshold $\epsilon$ (line 14). Following is the Pruning Strategy \ref{strategy: NBPS} (lines 16-21): After the entry of the first event of the last sub-episode of $\mathcal{Q}_e$ into $\mathcal{T}_a$ (i.e., after the search distance exceeds \textit{maxFirst}), all nodes in $\mathcal{T}_a$ except those containing the first event of $\mathcal{Q}_e$ and their descendant nodes are deactivated. Similarly, after the entry of the last event of the last sub-episode of $\mathcal{Q}_e$ into $\mathcal{T}_a$ (i.e., after the search distance exceeds \textit{maxLast}), all nodes in $\mathcal{T}_a$ except those containing all events of $\mathcal{Q}_e$ (nodes whose paths from the root node include the complete $\mathcal{Q}_e$) and their descendant nodes are deactivated. Deactivating nodes prevents them from participating in generating descendant nodes, i.e., it does not allow them to generate new rules.

The derivation of a new time list, \textit{timeList}, based on the end time set of episode $\alpha$ and the offset variable \textit{i} is as follows (lines 22-23). Each moment in \textit{timeList} is scanned to obtain the event set \textit{Event'} (lines 24-26). Subsequently, Pruning Strategy \ref{strategy: DBPS} is applied (lines 28-30): Nodes beyond the maximum search distance are not allowed to join $\mathcal{T}_a$ to generate descendants. Then, an internal loop related to $\mathcal{T}_a$ is executed (lines 31-47), significantly reducing resource consumption and improving operational efficiency \cite{ao2017mining}. To minimize the number of scans of the input time series, it is possible to judge whether the event \textit{u.event} can constitute the descendant node of another node based on the end time set \textit{u.tlist} of the node $u$.

In the internal loop, all existing nodes in $\mathcal{T}_a$ are traversed. Deactivated nodes need to be skipped in this internal loop according to Strategy \ref{strategy: NBPS} (lines 32-34). First, the time distance $d$ between the root node $r$ and the currently traversed node $w$ is calculated. Only when the current offset variable \textit{i} is greater than $d$, is there a possibility for the event \textit{u.event} to form a new descendant node of the node $w$. If the event \textit{u.event} occurs after the time interval ``\textit{i - d}" following \textit{w.event}, then all possible occurrence times are ``\textit{t + i - d}"(line 37). Since the node \textit{u.tlist} records the time intervals \textit{i} after the occurrence of the episode $\alpha$, all possible occurrence times of the event \textit{u.event} are recorded. Consequently,  the intersection $F$ of \textit{tmpList} and \textit{u.tlist} records all the times when \textit{u.event} occurs after \textit{w.event} after the time interval ``\textit{i - d}". If $F$ is greater than \textit{minSup}, the node $u'$ formed by \textit{u.event} and $F$ becomes a child node of $w$. Finally, Pruning Strategy \ref{strategy: LBPS} is applied: for mining rules whose length is less than the expected minimum rule length, they are not accepted. Instead, a new rule is generated.

\section{Experiments} \label{sec: experiments}

In this section, to evaluate the performance of the TaMIPER algorithm, we conducted a series of experiments. Specifically, we compared the performance of the TaMIPER algorithm against two others, MIPTRIE(DFS) and MIP-TRIE(PRU), using six real datasets. To meet the requirements of the target rules, we used post-processing techniques to filter out disqualified pattern rules. The analysis of algorithm performance was carried out comprehensively, considering aspects such as the number of candidate set patterns, runtime, the count of pruning strategies, other parameters, and memory consumption. All experiments were conducted on a computer with the Windows 10 operating system, equipped with an Intel(R) Core(TM) i7-10700F CPU and 16GB of RAM. All algorithms were implemented in Java. For reproducibility, both the source code and datasets are accessible on GitHub: \url{https://github.com/DSI-Lab1/TaMIPER}.

\subsection{Datasets}

We conducted the experiments on six real datasets, namely Retail, MSNBC, Kosarak, BMS1, BIKE, and SIGN. Retail represents sales data from a certain store's baskets; MSNBC, Kosarak, and BMS1 are datasets composed of website clickstreams; BIKE consists of location sequence data from shared bikes; and SIGN is a dataset related to sign language. The detailed information about the datasets used in the experiments is presented in Table \ref{table: datasets}. Here, $|E|$ denotes the number of events in the dataset, $|T|$ is the timespan of timestamps in the dataset, \textit{avg(L)} represents the average number of items per timestamp, and \textit{min(L)} and \textit{max(L)} denote the minimum and maximum values of items per timestamp, respectively.

\begin{table}[ht]
    \caption{Statistics of each dataset.}
    \label{table: datasets}
    \centering
    \small
    \resizebox{0.85\linewidth}{!}{
    \begin{tabular}{|c|c|c|c|c|c|}
        \hline
         \textbf{Dataset} & \textbf{$|E|$} & \textbf{$|T|$} & \textit{\textbf{avg(L)}} & \textit{\textbf{min(L)}} & \textit{\textbf{max(L)}} \\
         \hline
         \hline
         Retail & 16,470 & 87,997 & 10 & 1 & 50 \\ 
         \hline
         MSNBC & 17 & 31,790 & 5 & 2 & 17 \\
         \hline
         Kosarak & 41,270 & 966,819 & 5 & 1 & 50 \\
         \hline
         BMS1 & 497 & 595,35 & 2 & 1 & 50 \\ 
         \hline
         BIKE & 67 & 21,075 & 7 & 2 & 50 \\ 
         \hline
         SIGN & 267 & 340 & 41 & 18 & 50 \\ 
         \hline
    \end{tabular}}
\end{table}

For all experiments, we set the query episode $\mathcal{Q}_e$, the minimum support \textit{minSup}, the length threshold of the antecedent episode $\delta$, and the length threshold of the subsequent episode $\epsilon$ for each dataset, as shown in Table \ref{table: parameters}.

\begin{table}[ht]
    \centering
    \caption{Relevant parameters on each dataset.}
    \label{table: parameters}
    \begin{tabular}{|c|c|c|c|c|}
         \hline
         \textbf{Dataset} & $\mathcal{Q}_e$ & \textit{minSup} & $\delta$ & $\epsilon$ \\
         \hline
         \hline
         Retail & $<$38, 39$>$ & 2,000 & 4 & 5 \\ 
         \hline
         MSNBC & $<$9, 10$>$ & 2,000 & 4 & 5 \\ 
         \hline
         Kosarak & $<$10, 2$>$ & 8,000 & 3 & 5 \\ 
         \hline
         BMS1 & $<$33449, 33469$>$ & 10 & 4 & 10 \\ 
         \hline
         BIKE & $<$3005, 3005$>$ & 100 & 3 & 8 \\
         \hline
         SIGN & $<$1, 8$>$ & 60 & 3 & 6 \\ 
         \hline
    \end{tabular}
\end{table}

\begin{figure*}[ht]
    \centering
    \includegraphics[width=1\textwidth,trim=60 20 50 20,clip]{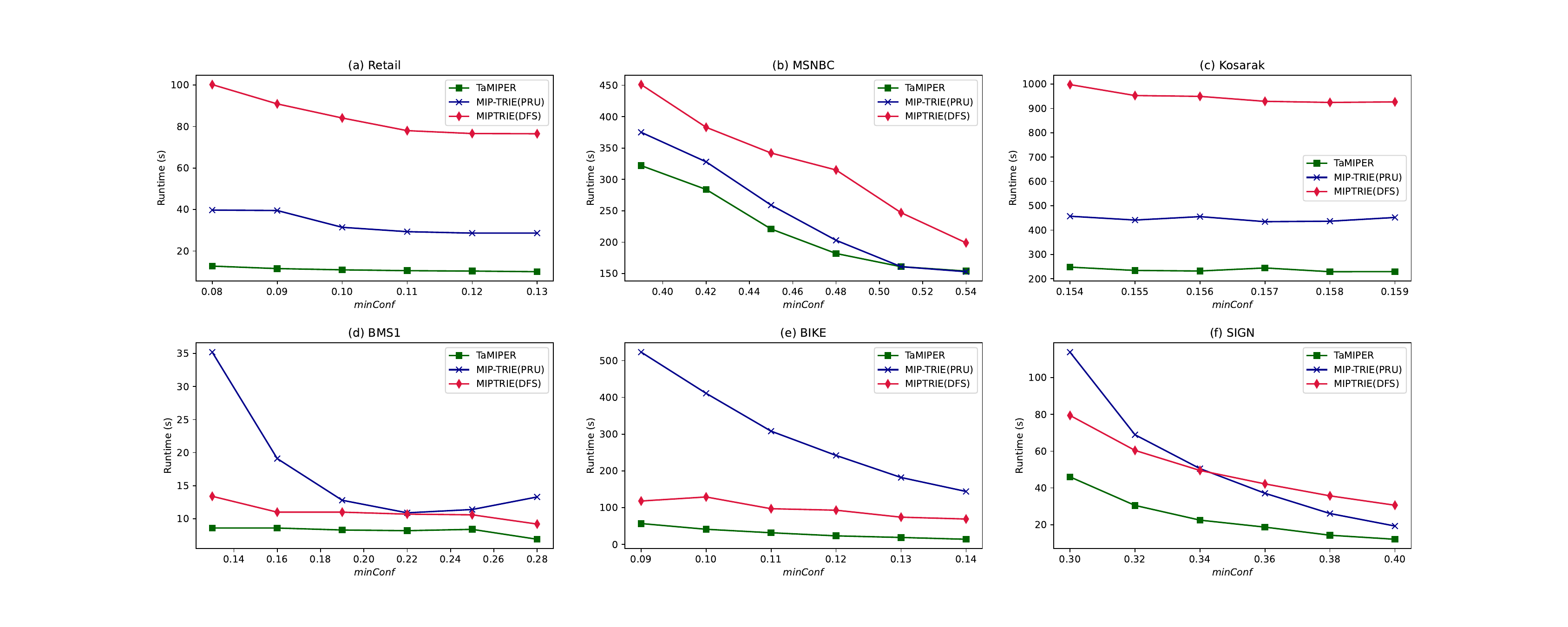}
    \caption{Runtime on each event dataset under different \textit{minconf}.}
    \label{fig:runtime}
\end{figure*}

\begin{figure*}[ht]
	\centering
	\includegraphics[width=1\textwidth,trim=60 20 50 20,clip]{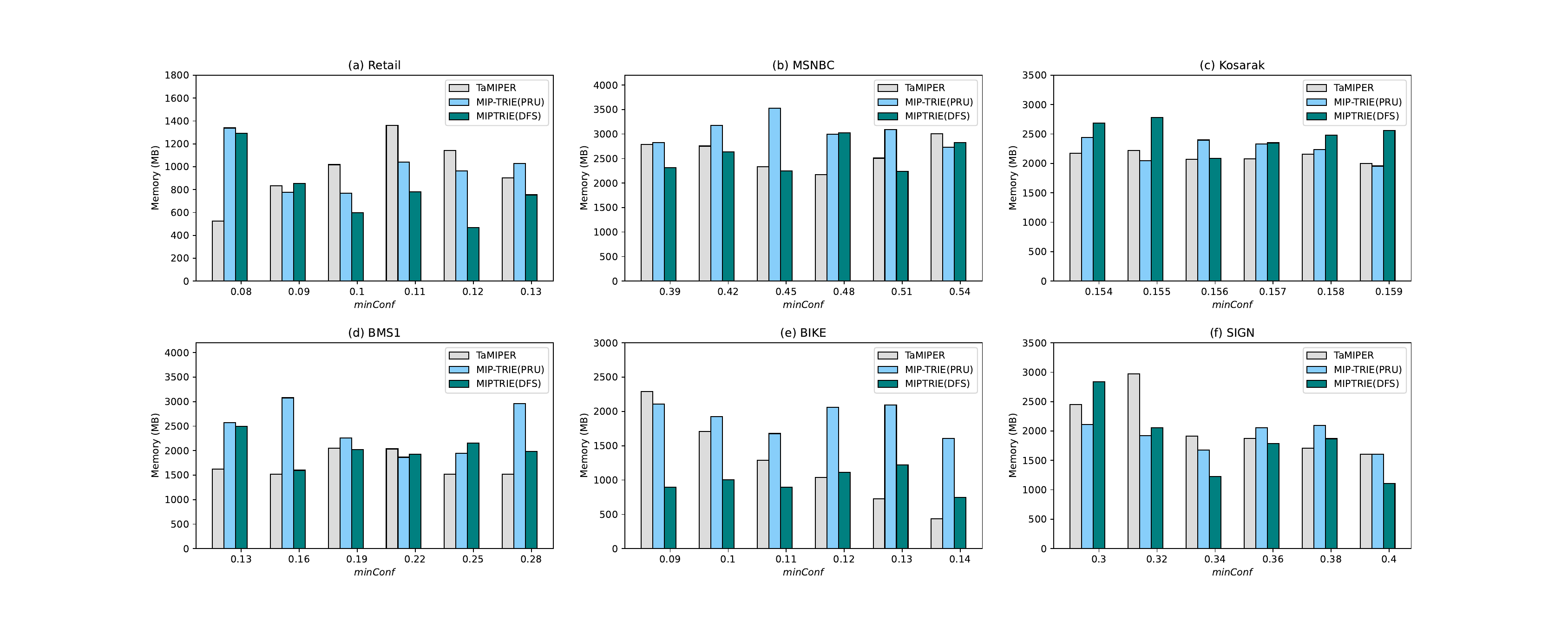}
	\caption{Memory consumption on each dataset under different \textit{minconf}.}
	\label{fig:memory}
\end{figure*}

\subsection{Runtime Evaluation}

The runtime required for mining episodes is a crucial metric for evaluating algorithm performance. Therefore, we conducted a series of experiments where the runtime varied with changes in \textit{minConf}. The important parameters related to the experiments are outlined in Table \ref{table: parameters}, and the settings for the subsequent experiments are the same. We compared and analyzed the runtime of three algorithms, including TaMIPER, MIPTRIE(DFS), and MIP-TRIE(PRU). Fig. \ref{fig:runtime} presents the results of these three algorithms running on the aforementioned six real datasets.

From Fig. \ref{fig:runtime}, it is clear that TaMIPER consistently outperforms MIPTRIE(DFS) and MIP-TRIE(PRU) in terms of runtime, regardless of the dataset used. This is because TaMIPER effectively utilizes the four pruning strategies proposed for targeted mining, setting it apart from the other two algorithms. While MIP-TRIE(PRU) displays slightly better performance than MIPTRIE(DFS), TaMIPER, which is based on MIP-TRIE(PRU), avoids generating a large number of invalid or meaningless episode rules during runtime. As a result, it significantly reduces the execution time of the algorithm. For example, as shown in Fig. \ref{fig:runtime}(a) for the Retail dataset, the runtime of MIP-TRIE(PRU) is 290\% of that of TaMIPER, while MIP-TRIE(DFS) takes 760\% longer. As the \textit{minConf} value gradually increases, the runtime difference between TaMIPER and the other two algorithms gradually narrows. This is because, with higher \textit{minConf} values, the proportion of effective TaPER rules among the discovered rules decreases, resulting in a reduced time difference. In other words, datasets with more regular patterns (a higher number of target episode rules) or lower minimum confidence thresholds highlight the pruning effectiveness of TaMIPER, leading to shorter runtimes compared to the other algorithms.

In summary, the TaMIPER algorithm demonstrates a significant time advantage over the two algorithms, MIPTRIE(DFS) and MIP-TRIE(PRU), in targeted mining precise episode rules.

\begin{table*}[ht]
    \renewcommand{\arraystretch}{1.35}
    \caption{The number of TaPERs and candidate rules under different \textit{minConf}.}
    \label{table:TaPERminconf}
    \centering
    \resizebox{0.85\linewidth}{!}{ 
    \begin{tabular}{|c|c|c|c|c|c|c|c|c|c|c|c|c|}
        \toprule\hline
         \textbf{Dataset} & \multicolumn{6}{c|}{\textbf{Retail}} & \multicolumn{6}{c|}{\textbf{MSNBC}} \\
         
         \hline
         \textit{minconf} & 0.08 & 0.09 & 0.10 & 0.11 & 0.12 & 0.13 & 0.39 & 0.42 & 0.45 & 0.48 & 0.51 & 0.54 \\
         \hline
         Candidate rules & 93,794 & 74,641 & 59,455 & 48,676 & 44,522 & 41,524 & 377,684 & 249,696 & 200,234 & 180,729 & 123,702 & 66,500 \\
         \hline
         TaPER rules & 3,144 & 2,795 & 2,215 & 1,128 & 537 & 139 & 96,991 & 42,401 & 33,083 & 33,051 & 22,781 & 116 \\
         \hline
         Percentage(\%)& 3.35 & 3.74 & 3.72 & 2.32 & 1.21 & 0.33 & 25.68 & 16.98 & 16.52 & 18.29 & 18.42 & 0.17 \\
         \hline
         
         \textbf{Dataset} & \multicolumn{6}{c|}{\textbf{Kosarak}} & \multicolumn{6}{c|}{\textbf{BMS1}} \\
         \hline
         \textit{minconf} & 0.154 & 0.155 & 0.156 & 0.157 & 0.158 & 0.159 & 0.13 & 0.16 & 0.19 & 0.22 & 0.25 & 0.28 \\
         \hline
         Candidate rules & 64,104 & 60,709 & 58,162 & 56,474 & 55,335 & 54,570 & 721,475 & 415,261 & 198,092 & 74,827 & 60,761 & 37,350 \\
         \hline
         TaPER rules & 3,145 & 2,173 & 1,428 & 907 & 596 & 342 & 721,475 & 415,261 & 198,092 & 74,827 & 60,761 & 37,350 \\
         \hline
         Percentage(\%)& 4.91 & 3.58 & 2.46 & 1.61 & 1.08 & 0.63 & 1.02 & 0.92 & 0.63 & 0.22 & 0.20 & 0.11 \\
         \hline
         
         \textbf{Dataset} & \multicolumn{6}{c|}{\textbf{BIKE}} & \multicolumn{6}{c|}{\textbf{SIGN}} \\
         \hline
         \textit{minconf} & 0.09 & 0.10 & 0.11 & 0.12 & 0.13 & 0.14 & 0.30 & 0.32 & 0.34 & 0.36 & 0.38 & 0.40 \\
         \hline
         Candidate rules & 1,561,938 & 1,284,849 & 1,078,835 & 916,042 & 777,428 & 661,460 & 939,571 & 628,378 & 447,000 & 332,232 & 249,378 & 193,403 \\
         \hline
         TaPER rules & 9,423 & 4,021 & 1,574 & 586 & 198 & 62 & 7,920 & 4,172 & 2,212 & 1,109 & 475 & 175 \\
         \hline
         Percentage(\%)& 0.60 & 0.31 & 0.15 & 0.06 & 0.03 & 0.01 & 0.84 & 0.66 & 0.50 & 0.33 & 0.19 & 0.09 \\
         \hline
         \bottomrule
    \end{tabular}}
\end{table*}

\subsection{Memory Evaluation}

Another important metric for evaluating algorithmic performance is the algorithm's memory usage. In order to demonstrate the effectiveness of the TaMIPER algorithm, we compared the memory consumption of three algorithms: TaMIPER, MIPTRIE(DFS), and MIP-TRIE(PRU). Fig. \ref{fig:memory} presents the performance of these three algorithms in terms of memory consumption on six real datasets.

From Fig. \ref{fig:memory}, we can see that the memory consumption of the TaMIPER algorithm is comparable to that of the other two algorithms overall, given the same dataset and \textit{minConf} conditions. In fact, in some cases, TaMIPER shows slightly superior memory consumption. This indicates that despite employing four pruning strategies to enhance runtime efficiency, TaMIPER does not add a significant burden on memory resources; in fact, it even manages to conserve certain memory resources. The memory utilization of the TaMIPER algorithm is influenced by different \textit{minConf} thresholds. Generally, there is a decreasing trend in memory consumption as \textit{minConf} increases. Alternatively, there may be an initial increase followed by a decrease in memory usage. The decrease in memory usage can be attributed to higher \textit{minConf} values, which result in fewer events meeting user requirements, leading to a smaller TaPER-tree size, a reduction in the number of rules, and consequently, lower memory usage. However, on the MSNBC and Kosarak datasets, the memory consumption of TaMIPER remains relatively stable. This stability could be due to minor variations in \textit{minConf} having a minimal impact on the construction of the TaPER-tree, resulting in insignificant changes in memory usage.

\begin{table*}[ht]
    \renewcommand{\arraystretch}{1.35}
    % \captionsetup{labelfont={color=red}}
    \caption{The number of rules under different pruning strategies. Note that the value of 0 indicates post-processing only (i.e., no pruning strategies), 1 indicates strategy \ref{strategy: PBPS} + post-processing, 2 indicates strategy \ref{strategy: PBPS} + strategy \ref{strategy: DBPS} + post-processing, etc.}
    \label{table: strategies}
    \centering
    \small
    \resizebox{0.9\linewidth}{!}{ 
    \begin{tabular}{|c|c|c|c|c|c|c|c|c|c|c|c|c|c|}
        \toprule\hline
         \multirow{2}{*}{\textbf{Dataset}} & \multirow{2}{*}{\textit{\textbf{minConf}}} & \multicolumn{5}{c|}{\textbf{Number of strategies used}} & \multirow{2}{*}{\textbf{Dataset}} & \multirow{2}{*}{\textit{\textbf{minConf}}} & \multicolumn{5}{c|}{\textbf{Number of strategies used}} \\ \cline{3-7} \cline{10-14} & & 0 & 1 & 2 & 3 & 4 &  &  & 0 & 1 & 2 & 3 & 4 \\
         \hline
         \multirow{3}{*}{Retail} & 0.07 & 106,466 & 50,793 & 49,325 & 32,720 & 15,514 & \multirow{3}{*}{MSNBC} & 0.39 & 377,684 & 361,193 & 354,514 & 334,375 & 195,692 \\ 
         \cline{2-7} \cline{9-14}
          & 0.08 & 93,794 & 39,576 & 38,267 & 25,748 & 11,897 & & 0.42 & 249,696 & 237,164 & 232,525 & 217,657 & 87,370 \\ 
         \cline{2-7} \cline{9-14}
          & 0.09 & 74,641 & 30,417 & 29,183 & 19,784 & 9,448 & & 0.45 & 200234 & 196214 & 192906 & 182981 & 66116\\ 
         \hline
         
         \multirow{3}{*}{Kosarak} & 0.154 & 64,104 & 41,742 & 39,360 & 32,090 & 3,921 & \multirow{3}{*}{BMS1} & 0.13 & 721,475 & 76,218 & 74,222 & 67,535 & 2,271 \\ 
         \cline{2-7} \cline{9-14}
          & 0.155	& 60,709 & 40,180 & 37,801 & 30,582 & 2,751 & & 0.16 & 415,261 & 41,501 & 40,419 & 36,789 & 7,641\\ 
         \cline{2-7} \cline{9-14}
          & 0.156 & 58,162 & 38,868 & 36,493 & 29,332 & 1,809 & & 0.19 & 198,092 & 17,044 & 16,675 & 15,575 & 14,332 \\ 
         \hline

        \multirow{3}{*}{BIKE} & 0.09 & 1,561,938 & 652,379 & 651,989 & 649,252 & 16,540 & \multirow{3}{*}{SIGN} & 0.30 & 939,571 & 789,959 & 783,035 & 752,613 & 334,118\\ 
         \cline{2-7} \cline{9-14}
          & 0.10 & 1,284,849 & 494,995 & 494,895 & 494,078 & 6,386 & & 0.32 & 628,378	& 513,568 & 511,155 & 498,798 & 163,377\\ 
         \cline{2-7} \cline{9-14}
          & 0.11 & 1,078,835 & 380,869 & 380,840 & 380,604 & 2,271 & & 0.34 & 447,000 & 353,258 & 352,505 & 347,620 & 80,080\\ 
         \hline
    \end{tabular}}
\end{table*}

\begin{figure*}[ht]
    \centering
    \includegraphics[width=1\textwidth,trim=50 20 50 20,clip]{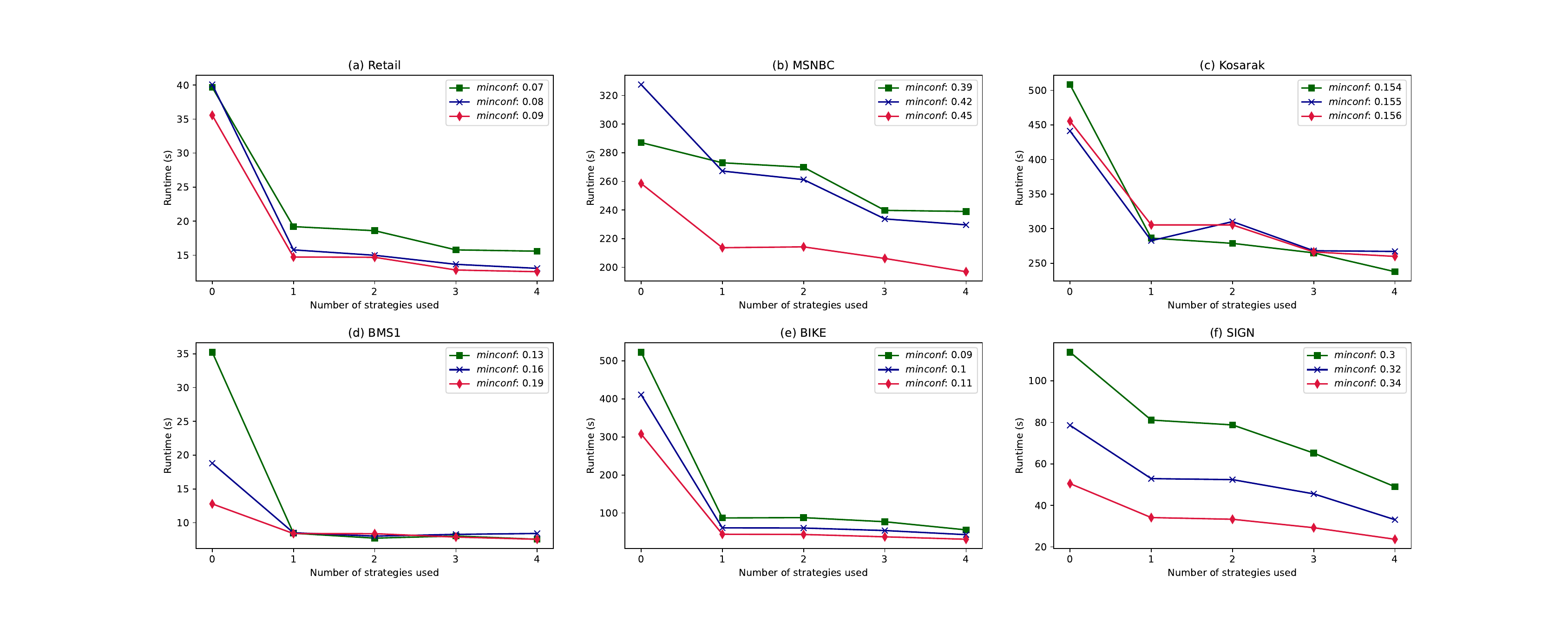}
    \caption{Runtime on each dataset with different strategies. On the x-axis, a value of 0 indicates post-processing only (i.e., no pruning strategies), 1 indicates strategy \ref{strategy: PBPS} + post-processing, 2 indicates strategy \ref{strategy: PBPS} + strategy \ref{strategy: DBPS} + post-processing, and so on.}
	\label{fig:strategyTime}
\end{figure*}

\subsection{Candidate Rule Evaluation}

The number of candidate rules serves as a crucial metric for measuring the reduction in ineffective computations. Table \ref{table:TaPERminconf} provides insights into the candidate rule counts, TaPER counts, and the proportion of TaPER to the total candidate rules discovered by TaMIPER across six different datasets. In this context, candidate rules pertain to those meeting user-specified threshold requirements such as \textit{minSup}, \textit{minConf}, \textit{delta}, \textit{epsilon}, etc., without considering whether they incorporate the query episode $\mathcal{Q}_e$. Meanwhile, TaPER denotes the target precise episode rules that satisfy all specified requirements.

From Table \ref{table:TaPERminconf}, it is evident that the quantity of TaPER rules is significantly less than that of candidate rules. For instance, in the MSNBC dataset, the highest proportion of TaPER is achieved, with an average of only 16.01\%. Conversely, in the BIKE dataset, the lowest proportion of TaPER is observed, with an average of 0.19\%. This discrepancy indicates that TaMIPER adeptly avoids generating a large number of invalid rules, thereby substantially reducing the computational load. It also underscores the effectiveness and meaningfulness of targeted mining. As \textit{minConf} increases, the proportion of TaPER rules noticeably decreases for each dataset. Combining this observation with the relationship between runtime and \textit{minConf} depicted in Fig. \ref{fig:runtime}, it becomes apparent that TaMIPER's performance is more favorable when the proportion of user-relevant TaPER in the candidate rules is higher—a scenario frequently encountered in practical applications.

\subsection{Pruning Strategy Comparison}

We further validate the effectiveness of the four pruning strategies in TaMIPER from two perspectives: the number of candidate rules and runtime. Table \ref{table: strategies} presents the effect of each pruning strategy on reducing the number of candidate rules. Firstly, comparing scenarios with no pruning strategy and utilizing all pruning strategies, the application of pruning strategies significantly reduces the number of rules on any dataset. For instance, on the Retail dataset with \textit{minConf} = 0.07, the four pruning strategies decrease the rule count from 106,466 to 15,514, which is roughly an 85\% reduction. Similar effects are observed on other datasets or at different confidence levels. Secondly, each pruning strategy has its own pruning capabilities. For instance, on the MSNBC dataset with \textit{minConf} = 0.39, strategy \ref{strategy: PBPS} reduces the rule count by 16,491 (377,684 – 361,193), strategy \ref{strategy: DBPS} by 6,679 (361,193 – 354,514), strategy \ref{strategy: NBPS} by 20,139, and strategy \ref{strategy: LBPS} by 138,683. Finally, it is worth noting that overall the strategy \ref{strategy: PBPS} exhibits the strongest pruning capability, attributed to its optimal utilization of positional information from query fragments. Following closely is the strategy \ref{strategy: LBPS}, which eliminates ineffective short rules that are challenging to prune with the first three strategies.

Fig. \ref{fig:strategyTime} presents the runtime requirements for different numbers of strategies across various datasets and confidence levels. The results demonstrate significant time savings achieved by implementing strategy \ref{strategy: PBPS}. Notably, this strategy performs exceptionally well on the BIKE dataset, reducing the runtime to approximately 15\% of the time required when relying solely on post-processing. While strategy \ref{strategy: DBPS} shows less pronounced improvement on the BMS1 dataset, which may be attributed to a lower average item count per timestamp in this dataset. In contrast, both strategy \ref{strategy: NBPS} and strategy \ref{strategy: LBPS} consistently contribute to time savings across all datasets, showcasing their stability and efficiency in enhancing TaMIPER's performance.

\subsection{Case Study of Rule Analysis}

The above experiments evaluated the performance of TaMIPER across various dimensions, including time, memory usage, and the generation of candidate rules. To fully demonstrate the algorithm's effectiveness, we analyzed the pattern rules mined by TaMIPER from a dataset of bike-sharing station locations. In this dataset, each item represents a bike-sharing station. Suppose we want to investigate the bike-sharing conditions at specific stations, such as station 3005 and station 3014. In this case, we set the query episode $\mathcal{Q}_e$ to $<$3005, 3014$>$, with other parameters shown in Table \ref{table: parameter}.

The results processed by our algorithm are shown in Table \ref{table: TaPERs}. A total of 11 TaPERS were extracted. Without setting the $\mathcal{Q}_e$ parameter, the baseline algorithms MIP-TRIE(PRU) and MIPTRIE(DFS) generate 565,625 rules, which contain excessive redundant information and fail to intuitively provide the bike origin information for stations 3005 and 3014. Detailed analysis of bike flow data allows us to extract source information for specific stations and analyze the riding demand and correlation between these stations and others, accurately identifying high-demand paths. For example, the first rule,``$<$3076, 3040$>$ $\overset{1}{\rightarrow}$ ($<$3005, 3014$>$, $<$2$>$)", indicates that bikes departing from station 3076, passing through station 3040, and finally reaching stations 3005 and 3014, have a support of 22 and a confidence of 18\%. Analyzing bike flow data in detail can also optimize bike scheduling, analyze user behavior, improve station layout, and enhance service quality. Overall, the rules mined by the TaMIPER algorithm are more refined and targeted, enabling the extraction of rules that are of greater interest to users.

\begin{table}[ht]
    \centering
    \renewcommand{\arraystretch}{1.5} % 调整行高
    \caption{Parameter settings of TaMIPER.}
    \label{table: parameter}
    \begin{tabular}{|c|c|c|c|c|c|}
         \hline
             \textbf{Dataset} & $\bm{\mathcal{Q}_e}$ & \textbf{\textit{minSup}} & \textbf{\textit{minConf}} & $\bm{\delta}$ & $\bm{\epsilon}$ \\
         \hline
             BIKE & $<$3005, 3014$>$ & 100 & 15\% & 3 & 8 \\
         \hline
    \end{tabular}
\end{table}

\begin{table}[ht]
    \centering
    \renewcommand{\arraystretch}{1.5} % 调整行高
    \caption{Valid TaPERs mined by TaMIPER.}
    \label{table: TaPERs}
    \resizebox{0.45\textwidth}{!}{
    \begin{tabular}{|c|c|c|}
         \hline
            \textbf{Rules} & \textbf{Support} $\downarrow$ & \textbf{Confidence} \\
         \hline
             $<$3076, 3040$>$ $\overset{1}{\rightarrow}$ ($<$3005, 3014$>$, $<$2$>$) & 22 & 18\% \\
             $<$3052, 3040$>$ $\overset{4}{\rightarrow}$ ($<$3005, 3014$>$, $<$4$>$) & 22 & 15\%\\
             $<$3069, 3014, 3027$>$ $\overset{2}{\rightarrow}$ ($<$3005, 3014$>$, $<$6$>$) & 18 & 17\%\\
             $<$3005, 3008, 3030$>$ $\overset{5}{\rightarrow}$ ($<$3005, 3014$>$, $<$1$>$) & 17 & 16\%\\
             $<$3034, 3064, 3005$>$ $\overset{5}{\rightarrow}$ ($<$3005, 3014$>$, $<$1$>$) & 17 & 15\%\\
             $<$3014, 3008, 3005$>$ $\overset{2}{\rightarrow}$ ($<$3005, 3014$>$, $<$3$>$) & 17 & 15\%\\
             $<$3014, 3008, 3005$>$ $\overset{4}{\rightarrow}$ ($<$3005, 3014$>$, $<$4$>$) & 17 & 15\%\\
             $<$3031, 3034, 3030$>$ $\overset{1}{\rightarrow}$ ($<$3005, 3014$>$, $<$5$>$) & 16 & 15\%\\
             $<$3074, 3069, 3031$>$ $\overset{5}{\rightarrow}$ ($<$3005, 3014$>$, $<$3$>$) & 16 & 15\%\\
             $<$3014, 3031, 3007$>$ $\overset{1}{\rightarrow}$ ($<$3005, 3014$>$, $<$6$>$) & 16 & 15\%\\
             $<$3031, 3075, 3014$>$ $\overset{3}{\rightarrow}$ ($<$3005, 3014$>$, $<$2$>$) & 15 & 15\%\\
         \hline
    \end{tabular}
    }
\end{table}

\subsection{Scalability}

\begin{figure}[ht]
    \centering
    \includegraphics[width=0.51\textwidth,trim=80 0 50 20,clip]{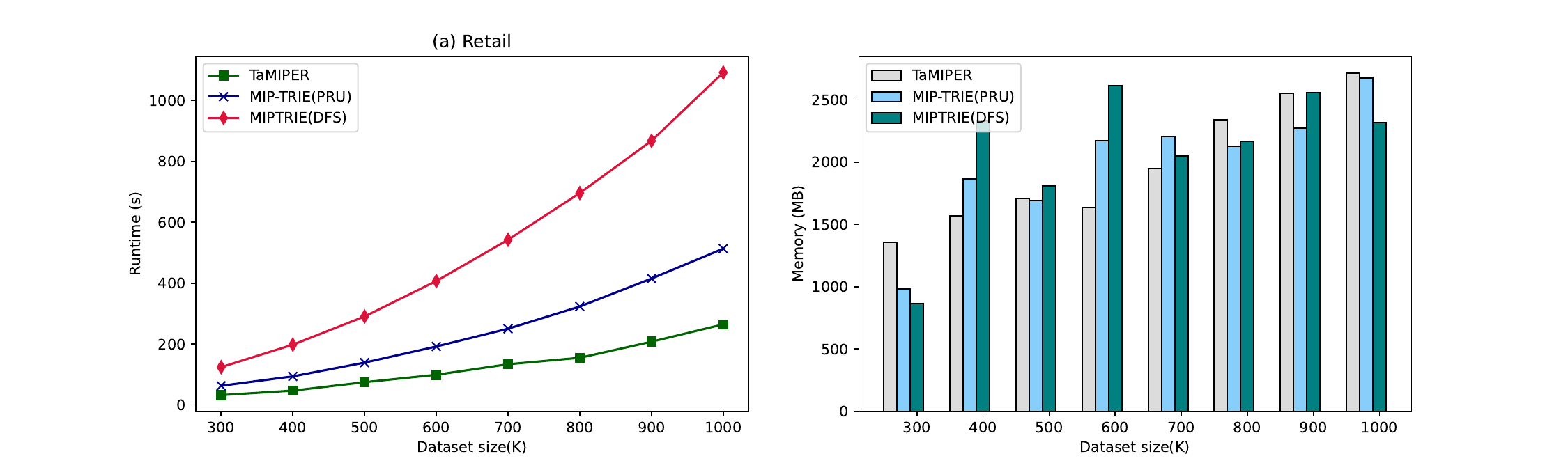}
    % \captionsetup{labelfont={color=red}}
    \caption{Runtime and memory usage under different dataset sizes.}
    \label{fig:Scalability}
\end{figure}

Finally, we evaluate the scalability of TaMIPER on eight synthetic datasets by comparing it with two baseline algorithms, MIP-TRIE(PRU) and MIPTRIE(DFS). The experiments utilize datasets ranging from 300K to 1000K timestamps, all maintaining the same characteristics. The evaluation metrics include execution time and memory usage. As shown in Fig. \ref{fig:Scalability}, the execution time of MIPTRIE(PRU) increases rapidly with the size of the dataset, and MIPTRIE(DFS) increases even more quickly. In contrast, TaMIPER exhibits a slow linear increase in execution time relative to the two baseline algorithms, indicating its efficiency in handling large datasets and its excellent scalability. In terms of memory usage, TaMIPER demonstrates stable performance, with memory consumption increasing slowly. This indicates that the algorithm does not require excessive additional memory to process large datasets. By comprehensively evaluating both metrics, it is evident that TaMIPER achieves slow increases in execution time and stable memory usage when processing large datasets. Therefore, TaMIPER has better scalability for processing big data.

\section{Conclusions and Future Work} \label{sec: conclusion}

In this paper, we introduce the definition of targeted precise-positioning episode rules and formulate the problem of targeted mining precise-positioning episode rules. Furthermore, we develop an algorithm called Targeted Mining Precision Episode Rules (TaMIPER) and optimize it using four strategies. Among these, Strategy \ref{strategy: PBPS} filters out the antecedents that do not meet the requirements using pre-extension time information.  Strategy \ref{strategy: DBPS} identifies nodes with a search distance greater than the maximum valid distance and excludes them during the generation of the TaPER-tree. Strategy \ref{strategy: NBPS} deactivates nodes that cannot grow into the superset of $\mathcal{Q}_e$.  Strategy \ref{strategy: LBPS} removes rules with a length smaller than the expected minimum target rule length. Finally, we conduct extensive experiments on various datasets to evaluate the performance of TaMIPER, and the results demonstrate its superiority.

In future research endeavors, we aim to delve deeper into topics related to targeted mining, as we believe it offers enhanced specificity and interest. Our plan involves designing sophisticated data structures tailored to different types of data requirements, considering various constraints such as non-overlapping occurrences and utility values. Finally, we aim to develop more robust algorithms and advanced pruning strategies to expand the application scope of targeted mining.

\bibliographystyle{IEEEtran}
\bibliography{TaMIPER.bib}

\end{document}